\documentclass[11pt]{article}
 \pdfoutput=1 
\usepackage{ucs}
\usepackage[utf8x]{inputenc}
\usepackage{fontenc}
\usepackage{graphicx}
\usepackage{float}
\usepackage{subfigure}
\usepackage{color}
\usepackage{amsmath}
\usepackage{hyperref}
\usepackage{cite}
\usepackage{url}
\title{\bf{The Reach of INO for Atmospheric Neutrino Oscillation Parameters}}
\author{Tarak Thakore$^1$\,\thanks{email: tarak@tifr.res.in}~,~~Anushree Ghosh$^2$\,\thanks{email: anushree@hri.res.in}~,~~ Sandhya Choubey$^2$\,\thanks{email: sandhya@hri.res.in}~,~~Amol Dighe$^1$\,\thanks{email: amol@tifr.res.in},\\\\
\\
 {\it $^1$Tata Institute of Fundamental Research, Colaba, Mumbai 400 005, India }\\\\
 {\it $^2$Harish-Chandra Research Institute, Chhatnag Road, Jhusi, Allahabad 211 019, India}\\\\
  }

\begin{document}

\maketitle
\renewcommand{\thefootnote}{\fnsymbol{footnote}}

\begin{abstract}

The India-based Neutrino Observatory (INO) will host a 50 kt magnetized iron calorimeter (ICAL@INO) for the study of atmospheric neutrinos. Using the detector resolutions and efficiencies obtained by the INO collaboration from a full-detector GEANT4-based simulation, we determine the reach of this experiment for the measurement of the atmospheric neutrino mixing parameters ($\sin^2 \theta_{23}$ and $|\Delta m_{32}^2 |$). We also explore the sensitivity of this experiment to the octant of $\theta_{23}$ , and its deviation from maximal mixing.

\end{abstract}

\section{Introduction}

Neutrino flavor oscillations have been established beyond any doubt by a series of 
outstanding results from 
atmospheric \cite{Wendell:2010md}, solar \cite{Aharmim:2011vm}, 
reactor \cite{Abe:2008aa,An:2012eh,Ahn:2012nd,Abe:2012tg} and accelerator \cite{Ahn:2006zza,Adamson:2012gt,Abe:2011ks} 
neutrino experiments. Neutrino flavor oscillations require neutrinos to  
be massive and mixed, and they have thus provided the first unambiguous hint for physics 
beyond the standard model of elementary particles. 
The neutrino mixing matrix, called $U_{\text{PMNS}}$ \cite{Pontecorvo:1967fh,Maki:1962mu},
can be parameterized in terms of three mixing angles $\theta_{12}$, $\theta_{13}$, $\theta_{23}$, and 
a charge-parity violating (CP) phase $\delta_{\text{CP}}$. In addition, if neutrinos were Majorana particles, we would 
have two Majorana CP phases $\alpha_1$ and $\alpha_2$ as well. The frequencies of neutrino oscillations are governed by two mass squared differences, $\Delta m_{21}^2$ and $\Delta m_{31}^2$, where we define 
$\Delta m_{ij}^2=m_i^2 - m_j^2$.
The so-called solar neutrino 
oscillation parameters $\theta_{12}$ and $\Delta m_{21}^2$ have been measured from 
the combined analysis of the KamLAND reactor $\overline{\nu}_e$ data and the  solar neutrino data. 
The so-called atmospheric neutrino 
oscillation parameters $\theta_{23}$ and $|\Delta m_{31}^2|$ 
are mostly constrained by the Super-Kamiokande (SK) atmospheric, and MINOS as well as T2K $\nu_\mu$ disappearance data. 
The third and the last mixing angle $\theta_{13}$ is the latest to be measured 
by a series of accelerator and reactor experiments. After decades of speculation on 
whether $\theta_{13}$ was zero,  data from these experiments have revealed that   
the value of $\theta_{13}$ is not only non-zero, it is 
in fact just below the previous upper bound  \cite{Apollonio:2002gd} from the Chooz experiment. 
Accelerator-based neutrino experiments 
T2K and MINOS have both observed $\nu_e$ appearance events 
from a beam of $\nu_\mu$ that indicates a non-zero value of $\theta_{13}$. 
The short baseline reactor neutrino experiments Daya Bay, RENO and Double Chooz have excluded $\theta_{13} = 0$ at 5.2$\sigma$, 4.9$\sigma$ and 3.1$\sigma$ respectively from $\overline{\nu}_e$ disappearance. 
Their best fit values are $\sin^2 2\theta_{13}=0.092\pm 0.016({\rm stat})\pm0.005({\rm syst})$ \cite{An:2012eh}, $\sin^2 2\theta_{13} = 0.113 \pm 0.013(\rm stat) \pm 0.019(\rm syst)$ \cite{Ahn:2012nd} and $\sin^2 2\theta_{13} = 0.109 \pm 0.030(\rm stat) \pm 0.025(\rm syst)$ \cite{Abe:2012tg}, respectively.
We summarize our current understanding of the neutrino oscillation parameters in Table \ref{tab_osc_param}.

 \begin{table}[H]

  \begin{center}
  \begin{tabular}{|c|c|c|c|c|}
     
  \hline
  Parameter & Best Fit Value & 3$\sigma$ Ranges \\
  \hline
  $\sin^2 \theta_{12}$ & 0.307 & 0.259-0.359\\
  \hline
  $\sin^2 \theta_{23}$ & 0.386 & 0.331-0.637 (NH)\\
			 & 0.392 & 0.335-0.663 (IH)\\
  \hline
  $\sin^2 \theta_{13}$ & 0.0241 & 0.0169-0.0313 (NH)\\
			  & 0.0244 & 0.0171-0.0315 (IH)\\
  \hline
  $\Delta m_{21}^2 \: (\rm eV^2)$ & 7.54 $\times \: 10^{-5}$ & 6.99-8.18 $\times \: 10^{-5}$\\
  \hline
  $|\Delta m_{31}^2| \: (\rm eV^2)$ & 2.43 $\times \: 10^{-3}$ & 2.19-2.62 $\times \: 10^{-3}$ (NH) \\
					  & 2.42 $\times \: 10^{-3}$ & 2.17-2.61 $\times \: 10^{-3}$ (IH) \\
  \hline
  
  \end{tabular}
  \end{center}

  \caption{A summary of the current values of the neutrino oscillation parameters. The values are taken from \cite{Fogli:2012ua}.
NH and IH indicate the limits applicable if the mass hierarchy is normal and inverted, respectively.}
  \label{tab_osc_param}

  \end{table}

While the above experiments continue to improve the precision on 
the mixing angle $\theta_{13}$, the focus has now shifted to the determination of the other unknown 
parameters in the neutrino sector. These include the CP violating phase $\delta_{\text{CP}}$ 
and the sign of $\Delta m^2_{31}$. Though the current global analyses \cite{Fogli:2012ua,Tortola:2012te} 
have started to provide some hints about the value of $\delta_{CP}$, these are still 
early days, and better data from dedicated experiments would probably be required before 
one could make any definitive statement on whether CP is violated in the lepton 
sector as well. The answer to this question would have far-reaching implications, as a positive answer would lend support to 
the idea of baryogenesis via leptogenesis during the early universe.

Determining the sign of $\Delta m^2_{31}$ is also of crucial importance, since its knowledge is essential for constructing 
the mass spectrum of neutrinos. This sign can be extracted via the detection of earth matter effects in atmospheric and 
accelerator-based neutrino beams. 
Proposed detectors aiming to observe earth matter effects 
in atmospheric neutrinos 
include the Iron CALorimeter (ICAL) at the India-based Neutrino Observatory (INO) 
\cite{Athar:2006yb}, megaton water Cerenkov detectors such as the 
Hyper-Kamiokande (HK) \cite{Abe:2011ts}, 
large liquid argon detectors \cite{Rubbia:2004nf,Akiri:2011dv} and a gigaton-class ice detector such as the  
Precision IceCube Next Generation Upgrade (PINGU) \cite{Koskinen:2011zz,Akhmedov:2012ah}. The 
accelerator beam neutrino experiment NO$\nu$A\cite{Ayres:2004js}, which will start operating soon, 
will also be sensitive to the mass hierarchy.

The other neutrino parameter that is yet to be measured is the 
absolute neutrino mass scale, on which currently we have only upper bounds from cosmological data \cite{Spergel:2003cb,Hannestad:2010kz,Lesgourgues:2012uu}, from neutrinoless double beta decay \cite{KlapdorKleingrothaus:2000sn,Ackerman:2011gz}, 
and from tritium beta decay experiments \cite{Kraus:2004zw,Wolf:2008hf}. 
Finally, the neutrinoless double beta decay experiments 
will also answer the most fundamental question -- whether neutrinos are Dirac or Majorana particles. 
If neutrinos are indeed Majorana, the neutrinoless double beta decay experiments 
might also have a chance to shed light on the Majorana phases 
in the far future.

INO is an experiment \cite{Athar:2006yb} proposed to study atmospheric neutrinos, employing the ICAL that can distinguish between $\mu^+$ and $\mu^-$, and will have good energy as well as direction resolution for muons. 
The magnetic field will enable it to distinguish
between neutrinos and anti-neutrinos, hence allowing identification of the neutrino mass hierarchy.
It will be located in the Bodi West Hills in the Theni district of Tamil Nadu in Southern India. The ICAL cavern will be located under 
a 1589 m high mountain peak. This peak provides a minimum rock cover of  about 1 km in all directions to reduce the cosmic muon background. 
The detector is designed to consist of 150 alternate layers of 5.6 cm thick iron plates and 
glass Resistive Plate Chambers (RPCs) 
stacked on top of each other. The total mass of the detector will be about 50 kt. The iron plates act as the 
target mass for neutrino interactions and the RPCs are the active detector elements that 
give the trajectory of the muons and hadrons passing through them. Iron will be magnetized to a field of about 1.3-1.5 Tesla. 

The main physics objective of ICAL is the determination of the neutrino mass hierarchy 
through the observation of earth matter effects in atmospheric neutrinos. 
In a recent analysis  \cite{Ghosh:2012px} performed by members of the INO collaboration, 
it was shown that data from this experiment along with that from the 
reactor and long baseline experiments T2K and NO$\nu$A could 
determine the neutrino mass hierarchy at the $2.2\sigma - 5.5\sigma$ C.L., depending on the true values of the parameters $\theta_{13}$, $\theta_{23}$ 
and $\delta_{\text{CP}}$, with 50 kt $\times$ 10 years of exposure.

In this paper, we explore in detail the potential for measuring the neutrino parameters $\theta_{23}$
 and $|\Delta m^2_{32}|$ in the ICAL@INO experiment using 
atmospheric neutrinos. The precision on both these parameters is expected to improve 
from data coming from the currently operating and soon-to-start experiments using accelerator-based 
neutrino beams (MINOS, T2K and NO$\nu$A) as well as atmospheric neutrinos 
(at Super-Kamiokande and IceCube Deep Core). Here we compare the potential of the 
ICAL@INO atmospheric neutrino experiment vis-a-vis the expected sensitivity of these 
complementary experiments. For our simulation of atmospheric neutrino events in ICAL@INO 
we use the software tailored and developed by the INO collaboration and present results 
obtained using the detector resolutions and efficiencies obtained from detailed simulations 
of the ICAL detector \cite{Reference_Muon,Reference_Hadron}. We restrict ourselves to using the information about the energy and direction of the muons produced in the detector. The results reported here are a part of the simulation and physics analysis being performed 
by the INO collaboration towards the full understanding of the capabilities of this experiment and a realistic study of its sensitivity reach. (Studies that explore the reach of a magnetized iron calorimeter for neutrino mixing have been performed in the past \cite{Samanta:2008af,Samanta:2010xm}, however this is the first time that we use the realistic ICAL detector simulations in the analysis.)

The paper is organized as follows. In section 2 we describe the detector simulation procedure for the ICAL detector, and the detector response for muons obtained from the GEANT4 simulations which are used in this work. 
In section 3, we outline our analysis procedure. In section 4, we discuss the results of our analysis, presenting the reach of the ICAL@INO for $\theta_{23}$ and $|\Delta m^2_{32}|$. Section 5 summarizes our findings.

\section{Simulation Framework and the Detector Response for Muons}

Neutrinos interact in the ICAL detector with nucleons and electrons, giving rise to leptons and hadrons.
As the outgoing lepton and hadrons propagate through the detector volume, they cross one or more RPC layers to give hit points in the RPCs. For every hit point, the $(x,y)$ position, time of hit and the RPC layer number are recorded. The information contained in the hit points is used to reconstruct the energy and direction of the particles. A muon passing through the ICAL detector gives hit points in several layers, which can be cleanly joined to form a  track. In contrast, a pion or other hadron passing though the detector typically gives rise to a hadron shower with multiple hit points in a single layer which cannot be joined to form a track. 
The ICAL detector is optimized primarily to measure the muon momentum with a good precision. The muon charge identification efficiency is also high.  The hadron energy and direction can be calibrated based on the number of hit points, and by combining the muon and hadron information, the neutrino momentum can be reconstructed, albeit with a coarser precision.

The detector simulation procedure is as follows. A GEANT4\cite{Agostinelli:2002hh,Allison:2006ve} -based detector simulation package for ICAL has been developed. The secondary particles are first propogated inside the detector using the GEANT4 code, taking into account the energy losses, multiple scattering, and magnetic field. This gives the hit positions of the particles in the active detectors, i.e. RPCs. Next, this output is digitized and used as an input for the reconstruction of the particle energy and direction. 

The reconstruction of the momentum of a particle is done in two steps. First, the topology of the digitized hit points is analyzed to find a possible muon track, and/or to find a hadron shower. Next, if the hit points are found to form a muon track, a Kalman Filter-based algorithm is used to reconstruct the initial energy, direction, and the vertex position of the muon. In case the hit points are identified as a hadron shower, the energy of the hadron is estimated by calibrating it against the number of hit points.

\begin{figure}[h]
 \centering
\subfigure[Muon energy Resolution]{
 \includegraphics[scale=0.37,keepaspectratio=true]{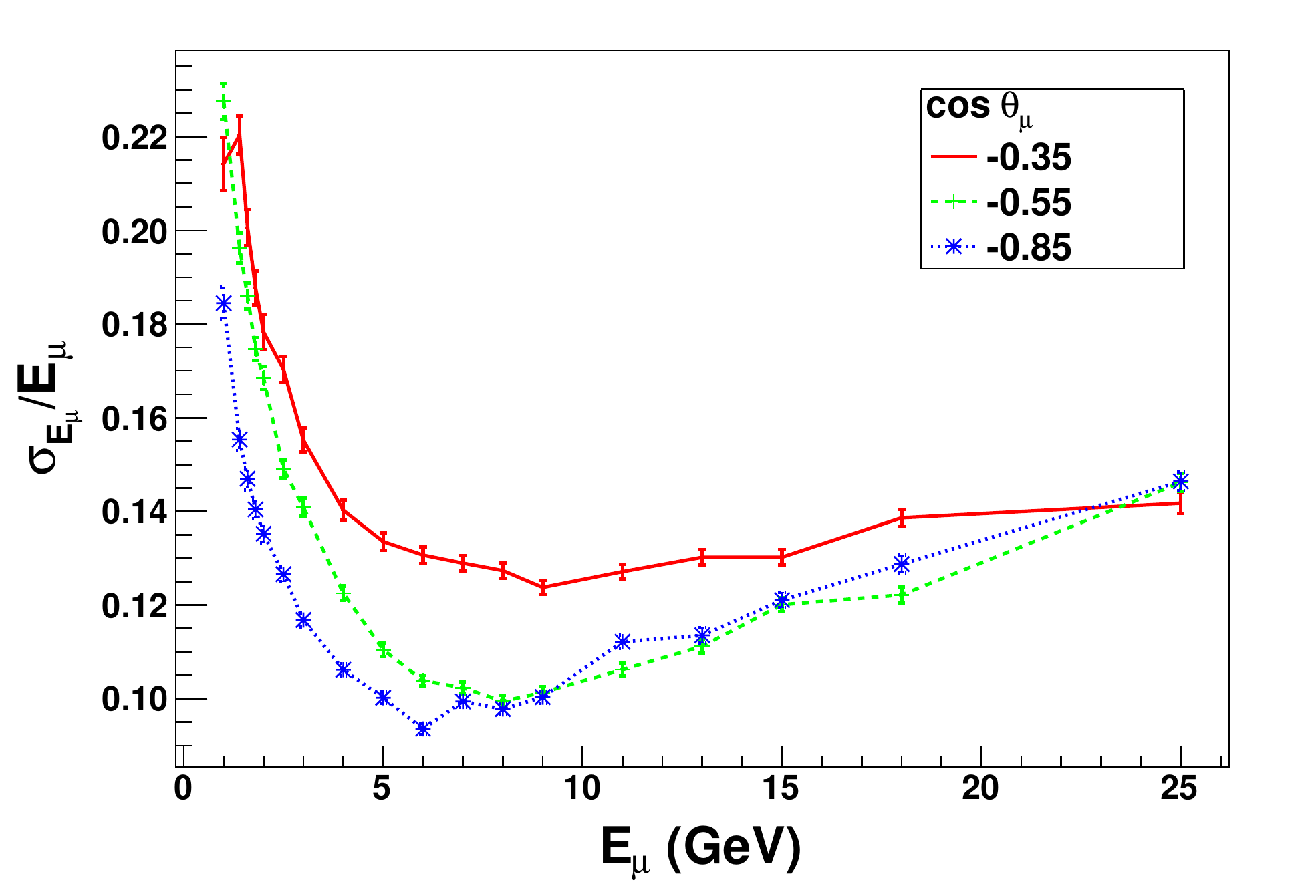}
 \label{fig:fig_Muon_En_Res}
}\quad
\subfigure[cos $\theta_\mu$ Resolution]{
 \includegraphics[scale=0.37,keepaspectratio=true]{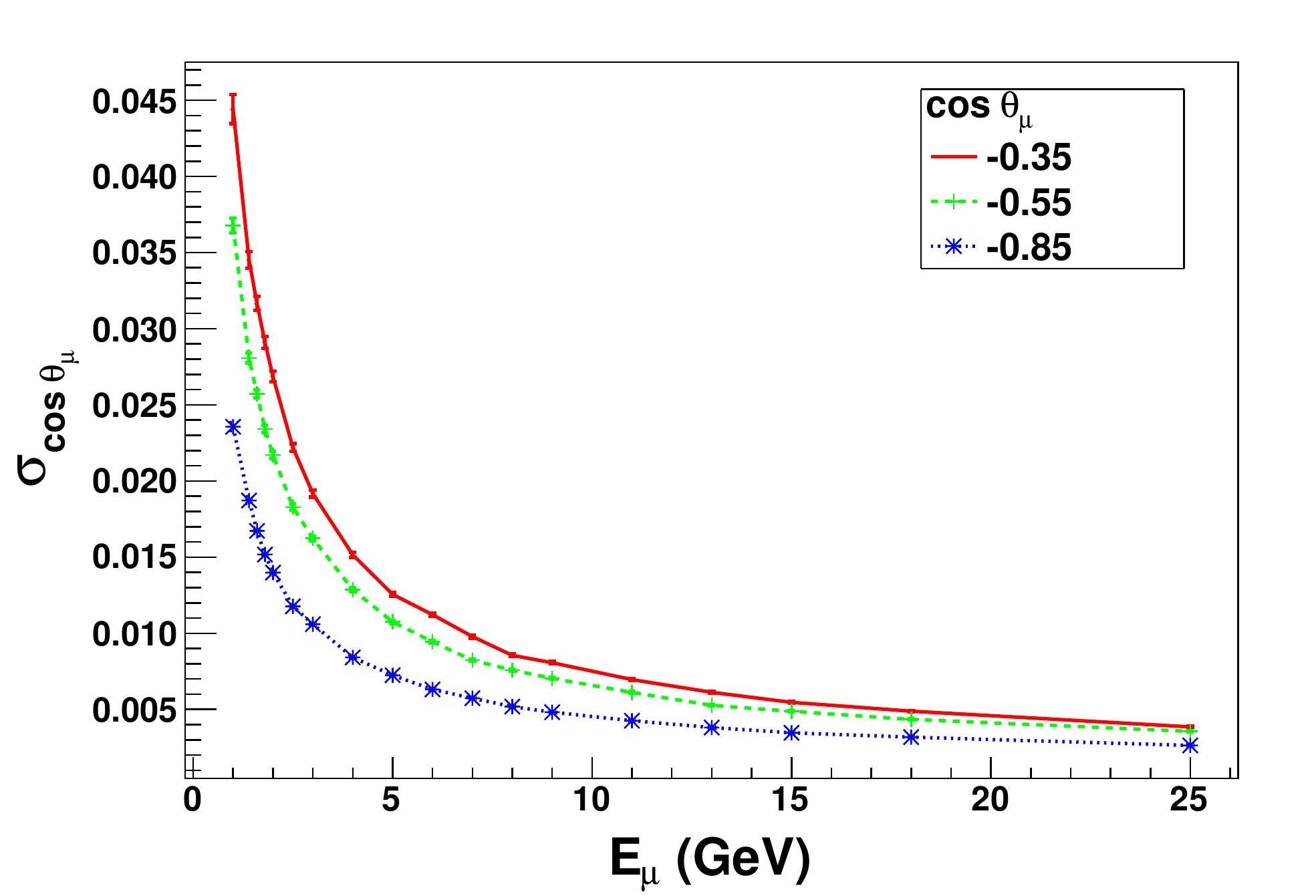}
 \label{fig:fig_Muon_CT_Res}
}
\subfigure[Reconstruction efficiency for muons]{
 \includegraphics[scale=0.37,keepaspectratio=true]{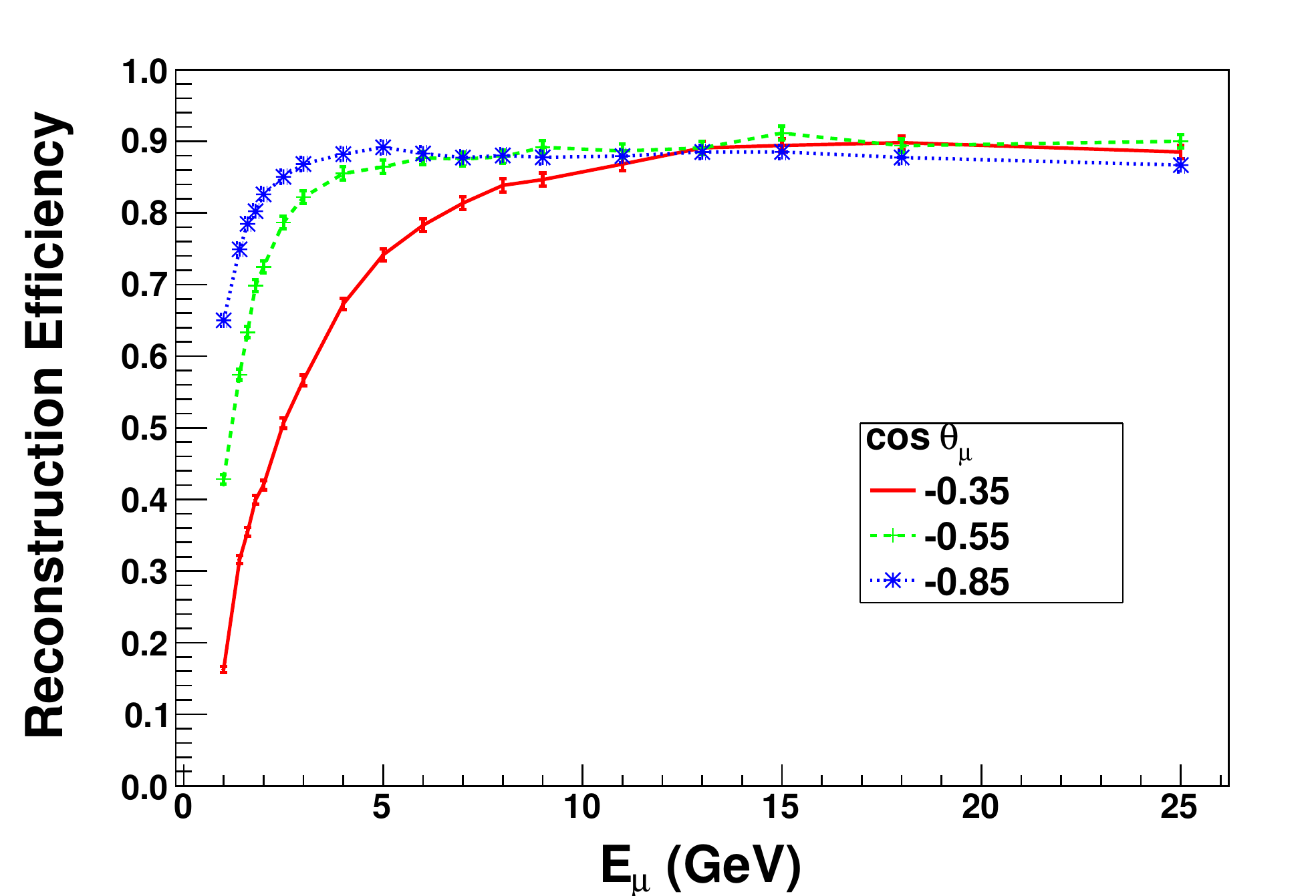}
 \label{fig:fig_Muon_Rec_Eff}
}\quad
\subfigure[Charge identification efficiency for muons]{
 \includegraphics[scale=0.37,keepaspectratio=true]{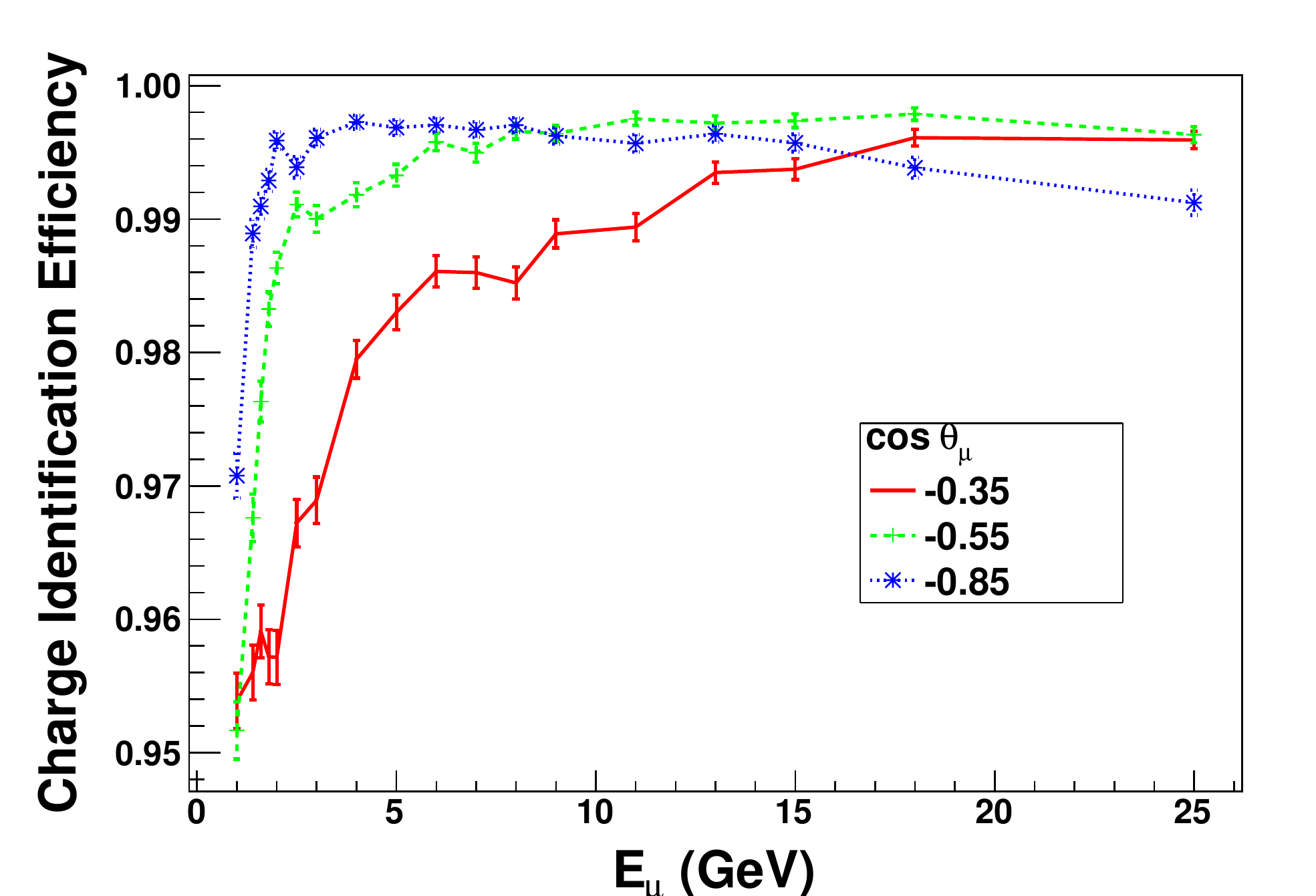}
 \label{fig:fig_Muon_CID_Eff}
}
 \caption{The energy resolution (a), $\cos \theta_\mu$ resolution (b), the reconstruction efficiency (c), and the charge identification efficiencies (d), 
 for $\mu^-$ as a function of the true muon energy and for three cases of true muon zenith angle. The red, green and blue lines are for the zenith angle bins with $\cos \theta_\mu = -0.35 \pm 0.05, -0.55 \pm 0.05, -0.85 \pm 0.05$, respectively.}
 \label{fig:fig_Muon_Recon}
\end{figure}

The muon response of the ICAL detector is parametrized in terms of the energy resolution ($\sigma_E$), direction resolution ($\sigma_{\cos \theta}$), reconstruction efficiency ($\epsilon_R$) and the charge identification efficiency ($\epsilon_C$). All these quantities are estimated as functions of the true muon energy ($E_\mu$) and direction ($\cos \theta_\mu$), for $\mu^-$ and $\mu^+$. Here $\theta_\mu$ is the zenith angle for the muon. Figure \ref{fig:fig_Muon_Recon} shows the muon energy and direction resolutions, as well as the reconstruction and charge identification (CID) efficiencies for $\mu^-$, for some specific muon zenith angles. The numbers for $\mu^+$ are virtually identical. We note here that the reconstructed energy distribution of the muon is fitted with the Normal distribution function for $E_\mu \ge$ 1 GeV, whereas for $E_\mu<$ 1 GeV it is fitted to the Landau distribution function. In the present work, we have carried out the oscillation analysis with muon energy and direction ($E_\mu$, $\
cos \theta_\mu$) as observables using simulated data. 

The analysis in this paper is based only on the energy, direction, and charge of the muon, for the charged-current
(CC) $\nu_\mu$ events. The inclusion of information on hadron energy and direction 
is beyond the scope of this paper.
The detector response to muons used in this work comes from the first set of 
simulations done with the ICAL code. These simulations are ongoing and are 
being fine-tuned. Hence the values of the resolution functions and efficiencies are
likely to evolve along with the simulations.

\section{Oscillation Analysis Procedure}

The physics analysis of atmospheric neutrino events requires simulations which can be broadly 
classified into four steps : (i) neutrino event generation, (ii) inclusion of the oscillation effects, 
(iii) folding in the detector response, and (iv) the $\chi^2$ analysis. 
These steps are described in detail in the following subsections.

\subsection{Event Generation}
We use the neutrino event generator NUANCE (version 3.5) \cite{Casper:2002sd} to generate the neutrino interactions. The atmospheric neutrino fluxes provided by Honda et al. \cite{Honda:2011nf} at the Super Kamiokande location are used.\footnote{The neutrino fluxes at Theni have not yet been incorporated in the NUANCE event generator.} The ICAL detector composition and geometry are specified as an input. NUANCE incorporates the differential cross section for charged-current (CC) and neutral-current (NC) interactions for all nuclear constituents of the materials used in the detector, for all neutrino flavors, and for all possible interaction processes. The quasi-elastic (QE), resonance (RS), deep inelastic (DIS), coherent (CO) and diffractive (DF) scattering processes are all included. The neutrino fluxes are then multiplied with the interaction cross-sections to calculate the event rates for all scattering processes and possible target nuclei. Event kinematics are generated based on the differential cross-
sections. 

The NUANCE output consists of the 4-momentum ($p^\mu$) of the initial, intermediate and the final state particles for each event.
To reduce the Monte Carlo (MC) fluctuations in the number of events given by NUANCE, we generate a very large 
number of neutrino interactions (an exposure of 50 kt $\times$ 1000 years) and scale it down to the desired 
exposure for the $\chi^2$ analysis.

\subsection{Inclusion of Oscillations}

The total number of $\nu_\mu$ events coming from the $\nu_\mu \rightarrow \nu_\mu$ 
and the $\nu_e \rightarrow \nu_\mu$ channels is given as 
\begin{equation} \label{eq_event_rate}
 \frac{d^2N}{dE_\nu \: d(\cos \theta_\nu)} = 
 N_T \times N_D \times \sigma_{\nu_\mu} \times 
 \left[P_{\mu\mu}  \frac{d^2\Phi_{\nu_\mu}}{dE_\nu \: d(\cos \theta_\nu)}
 + P_{e\mu} \frac{d^2\Phi_{\nu_e}}{dE_\nu \: d(\cos \theta_\nu)}
 \right] \,,
\end{equation}
where $N_T$ is the exposure time and $N_D$ is the number of targets 
in the detector. Here $\Phi_{\nu_\mu}$ and $\Phi_{\nu_e}$ are the fluxes of 
$\nu_\mu$ and $\nu_e$ respectively, and $P_{\alpha\beta}$ is the $\nu_\alpha \rightarrow \nu_\beta$ 
oscillation probability. Since an exposure of 1000 years has to be taken to reduce the MC fluctuations, 
generating events using NUANCE for each set of oscillation parameters is extremely time consuming and 
practically impossible within a reasonable time frame. Therefore, we generate events for 1000 years of 
exposure using NUANCE only once and employ the event re-weighting method described below to take care of neutrino oscillations
on this event sample for any given set of oscillation parameters. 

To get the number of $\mu^-$ events from $\nu_\mu$ that have survived oscillations, 
we generate the $\nu_\mu$ interactions from NUANCE using the un-oscillated $\nu_\mu$ flux in the 
neutrino energy 
range $0.5-100$ GeV and all neutrino zenith angles $\theta_\nu$. 
For a given cos $\theta_\nu$, the path travelled between the production 
point and the detector is
  \begin{equation}\label{eq_pathlength}
   L = \sqrt{(R \: + \: L_0)^2 \: - \: (R \: \sin \theta_\nu)^2} \: - \: R \: \cos \theta_\nu \,,
  \end{equation}
where $R$ is the radius of the earth (6378 Km) and $L_0$ is the average height of the 
atmospheric neutrino production, taken here to be 15 km. 
The survival probability $P_{\mu\mu}$ for the neutrino energy $E_\nu$ and zenith angle $\theta_\nu$ for a given event 
is calculated using the oscillation probability code included in NUANCE \cite{Casper:2002sd}. 
The details of this oscillation probability calculation are as described in \cite{Barger:1980tf}. 
We next impose the event re-weighting algorithm as follows. 
To decide whether an un-oscillated $\nu_\mu$ event survives the oscillations to be detected as $\nu_\mu$, a uniform random number $r$ is generated between 0 and 1. If $P_{\mu e} \leq r < P_{\mu e} + P_{\mu\mu}$, we keep this event as a $\nu_\mu$ event. Otherwise this $\nu_\mu$ is considered to have oscillated into a different flavor. 

Similarly, to get the $\mu^-$ events from atmospheric $\nu_e$ 
converted to $\nu_\mu$ due to flavor oscillations, we generate the 
$\nu_\mu$ interactions from NUANCE using the un-oscillated $\nu_e$ flux in the 
neutrino energy 
range $0.5-100$ GeV and all neutrino zenith angles $\theta_\nu$. This then has to be 
folded in with the conversion probability $P_{e\mu}$. For that we again use the event 
re-weighting algorithm where for each event a random number $r$ is generated between 0 and 1. 
If $r < P_{e \mu}$ we label the event as a $\nu_\mu$ event, otherwise we discard it.

The $\nu_\mu$ events from $\nu_\mu \rightarrow \nu_\mu$ as well as $\nu_e \rightarrow \nu_\mu$ 
oscillation channels are added to form the $\mu^-$ event sample. Similarly, we form the $\mu^+$ event 
sample as well. 
Next, we read the muon momentum of all the events that we have collected after oscillations and bin them according to energy ($E_\mu$) and direction ($\cos \theta_\mu$) of the muons. We bin the data in the energy range  $E_\mu$ = 0.5 - 15.5 GeV (300 bins) and 
$\cos \theta_\mu$ in the range $-1$ to $+1$ (20 bins) because our analysis threshold is 0.8 GeV and the muon reconstruction efficiency is very low below 0.5 GeV. We keep track of the numbers of $\mu^-$ and $\mu^+$ events separately. 
At this stage, we have the distribution of muon events in terms of their ``true'' energy ($E_\mu$) and the cosine of the zenith angle ($\cos \theta_\mu$). 

\begin{figure}[H]
 \centering
 \includegraphics[scale=0.6,keepaspectratio=true]{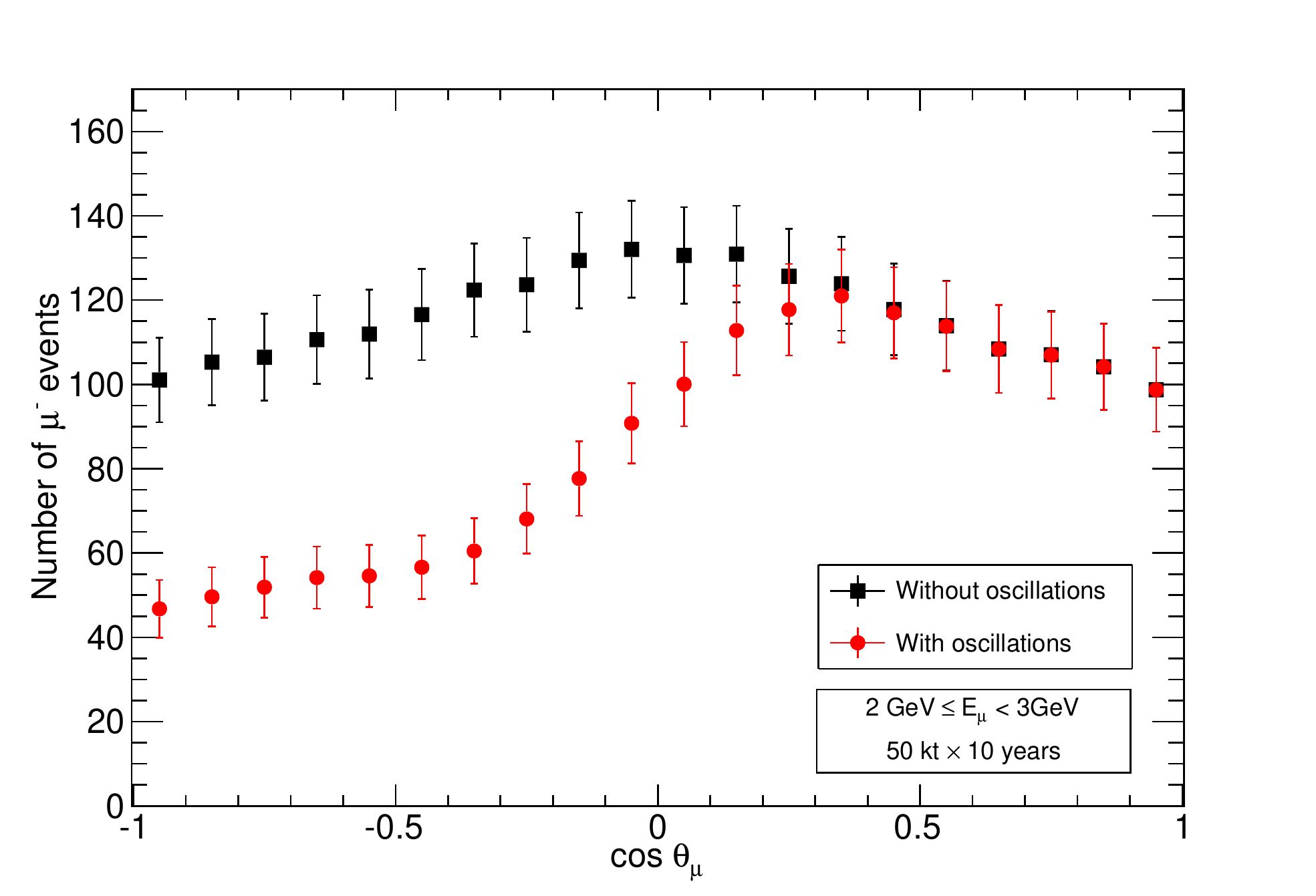}
 \caption{Zenith angle distribution of $\mu^-$ events for the bin 2 GeV $\leq E_\mu<$ 3 GeV, without and with flavor oscillations. The detector efficiencies have not been included here. The error bars shown here are statistical.}
 \label{fig:fig_MC_Osc}
\end{figure}

Figure~\ref{fig:fig_MC_Osc} shows the zenith 
angle distribution of $\mu^-$ events in the energy bin $2-3$ GeV before and after invoking oscillations using the re-weighting algorithm described above. We use the oscillation  parameters described in Table~\ref{tab_osc_param_input} and take the exposure to be 50 kt $\times$ 10 years. \\

\begin{table}[h]
\begin{center}
\begin{tabular}{|l|l|l|l|l|l|l|l|}
\hline
Parameter & $\sin^2 2\theta_{12}$ & $\sin^2 2\theta_{23}$ & $\sin^2 2\theta_{13}$ &  $\Delta m_{21}^2 \: (\rm eV^2)$ & $|\Delta m_{32}^2| \: (\rm eV^2)$ &  $\delta_{\text{CP}}$ & Hierarchy\\
\hline
True Value & 0.86 & 1.0 & 0.113 & 7.6 $\times \: 10^{-5}$ & 2.424 $\times \: 10^{-3}$ & 0.0 & Normal \\
\hline
\end{tabular}
\end{center}
\caption{True values of the neutrino oscillation parameters used in this paper}
\label{tab_osc_param_input}
\end{table}


\subsection{Folding in the Detector Response}

Next, we fold in the detector response \cite{Reference_Muon} to obtain the measured distribution of muons. 
We apply the reconstruction efficiency ($\epsilon_{R-}$) for $\mu^-$ by multiplying the 
number of events in a given true energy ($E_\mu$) and true zenith angle 
($\cos \theta_\mu$) bin with the corresponding reconstruction efficiency:
  \begin{equation}
    N_{\mu-}(E_\mu,\cos \theta_\mu) = \epsilon_{R-}(E_\mu,\cos \theta_\mu) \times N_{\mu-}^{\rm true}(E_\mu,\cos \theta_\mu) \,,
    \label{eq:evrecon}
  \end{equation}
where $N_{\mu-}^{\rm true}$ is the number of $\mu^-$ events in a given ($E_\mu$, $\cos \theta_\mu$) bin.
Exactly the same procedure is used for determining the $\mu^+$ events.
The CID efficiency ($\epsilon_{C-}$ for $\mu^-$ and $\epsilon_{C+}$ for $\mu^+$ event sample) is next applied as follows:
\begin{equation}
 N_{\mu^-}^C = \epsilon_{C-} \times N_{\mu^-} + (1 - \epsilon_{C+}) \times N_{\mu^+} \,,
  \label{eq:evcid}
  \end{equation}
where $N_{\mu^-}$ and  $N_{\mu^+}$ are the number of $\mu^-$ and $\mu^+$ events, respectively,
given by Eq.~(\ref{eq:evrecon}). Now $N_{\mu^-}^C$ is the number of $\mu^-$ events after taking care of the CID efficiency. All the quantities appearing in Eq.~(\ref{eq:evcid}) are functions of $E_\mu$ and $\cos \theta_\mu$.

\begin{figure}
 \centering
 \includegraphics[scale=0.6,keepaspectratio=true]{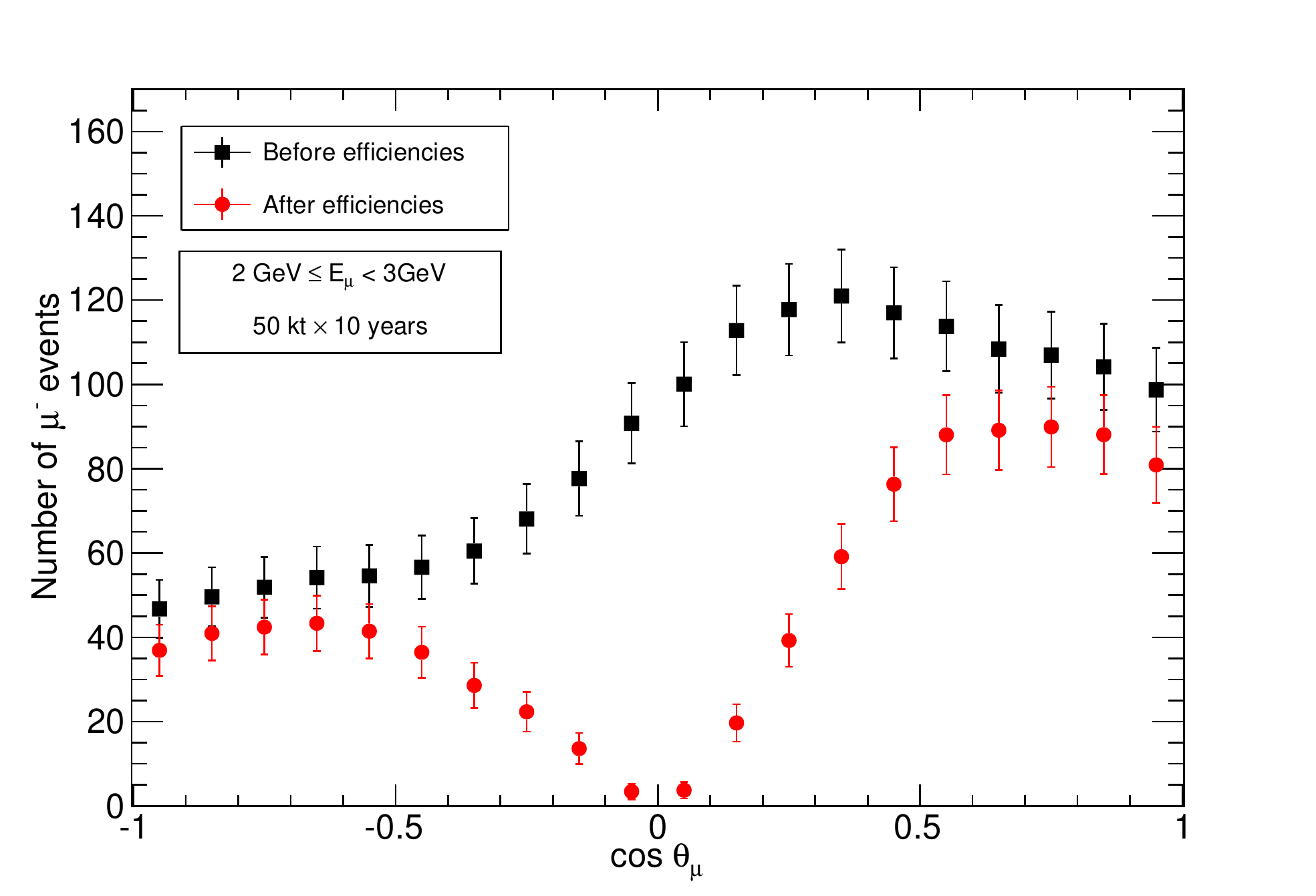}
 \caption{Zenith angle distribution of oscillated $\mu^-$ events for the bin 2 GeV $\leq E_\mu<$ 3 GeV, after taking into account detector efficiencies. The error bars shown here are statistical.}
 \label{fig:fig_event_distribution_with_Eff}
\end{figure}

Figure~\ref{fig:fig_event_distribution_with_Eff} shows the zenith angle distribution of events obtained before and 
after applying the reconstruction and CID efficiencies. Compared to Fig. \ref{fig:fig_MC_Osc}, one can notice that the number of events fall sharply for the almost horizontal ($\cos \theta_\mu \approx 0$) bins. This can be understood by noting from Figure \ref{fig:fig_Muon_Rec_Eff}
that the reconstruction efficiency for muons falls as we go to more horizontal bins. 
The reason for this is the ICAL geometry where iron slabs and RPCs are stacked horizontally.
Hence muons that are close to the horizontal direction cross fewer RPCs and mostly 
travel in the passive iron part of the detector. Reconstructing these tracks therefore becomes 
extremely difficult and the reconstruction efficiency of the detector falls.
The CID efficiency also falls for the more horizontal bins and the result is that there 
are hardly any events for bins with $-0.2 \leq \cos \theta_\mu < 0.2$. 

Finally, the muon resolutions $\sigma_E$ and $\sigma_{\cos \theta}$ are applied as follows:
\begin{equation}\label{eq_det_res}
 (N_{\mu^-}^D)_{ij} = \sum_k \sum_l N_{\mu^-}^C(E_\mu^k,\cos \theta_\mu^l) \,\,K_i^k(E_\mu^k) \,\,
 M_j^l(\cos \theta_\mu^l) \,,
\end{equation}
where $(N_{\mu^-}^D)_{ij}$ denotes the number of muon events in the $i^{th}$ $E$-bin and the $j^{th}$ $\cos \theta$-bin after applying the energy and 
angle resolutions. Here $E$ and $\cos \theta$ are the measured muon energy and zenith angle. 
The summation is over the true energy bin $k$ and true zenith angle bin $l$, with $E_\mu^k$ and $\cos \theta_{\mu}^l$ being the central values of the $k^{\rm th}$ true muon energy and $l^{\rm th}$ true muon zenith angle bin. The quantities  
$K_i^k$ and $M_j^l$ are the integrals of the detector resolution functions over the bins of $E$ and $\cos \theta$, the measured energy and direction of the muon, respectively. These are evaluated as:

\begin{equation}\label{eq_det_int_K}
 K_i^k (E_\mu^k) = \int_{E_{L_i}}^{E_{H_i}} dE \frac{1}{\sqrt{2\pi} \sigma_{E_\mu^k}} 
\exp \left( {- \frac{(E_\mu^k - E)^2}{2 \sigma_{E_\mu^k}^2} } \right) \,,
\end{equation}
and
\begin{equation}\label{eq_det_int_M}
M_j^l (\cos \theta_\mu^l) = \int_{\cos \theta_{L_j}}^{\cos \theta_{H_j}} d\cos  \theta\frac{1}{\sqrt{2\pi} 
\sigma_{\cos\theta_\mu^l}} 
\exp \left (  - \frac{(\cos \theta_\mu^l - \cos \theta)^2}{2 \sigma_{\cos\theta_\mu^l}^2} \right ) \,,
\end{equation} 
where $\sigma_{E_\mu^k}$ and $\sigma_{\cos\theta_\mu^l}$ are the energy 
and zenith angle resolutions, respectively, in these bins. 
We perform the integrations between the lower and upper boundaries of the measured energy 
($E_{L_i}$ and $E_{H_i}$) and the measured zenith angle ($\cos \theta_{L_j}$ and $\cos \theta_{H_j}$). 
For the extreme $\cos \theta$ bins, the bins are taken to be ($-\infty$, -0.9) and [0.9, +$\infty$) while integrating, and the events are assigned to the bins [-1, -0.9] and [0.9, 1], respectively. This ensures that no event is lost to the unphysical region and the total number of events does not change after
applying the angular resolution.
For $E_\mu^k <1$ GeV, the integrand in Eq.~(\ref{eq_det_int_K}) is replaced with the Landau distribution function, as the reconstructed energy distribution obtained from ICAL simulations \cite{Reference_Muon} 
is specified in terms of this function.

\begin{figure}[H]
 \centering
 \includegraphics[scale=0.6,keepaspectratio=true]{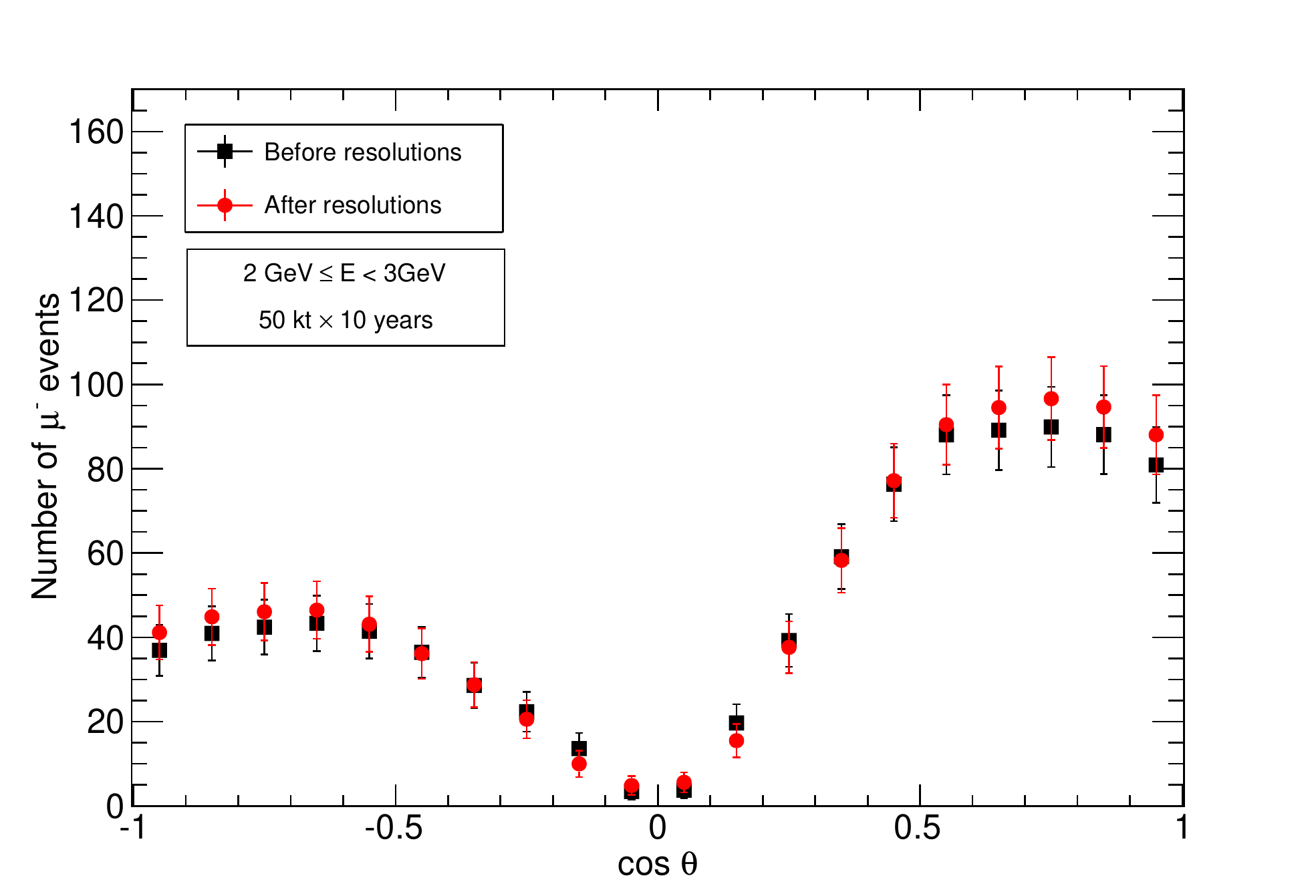}
 \caption{Zenith angle distribution of $\mu^-$ events for the bin 2 GeV $\leq E_\mu<$ 3 GeV before and after including energy and zenith angle resolution function. Here $E$ and $\theta$ are the measured energy and measured zenith angle, respectively. The error bars shown here are statistical.}
 \label{fig:fig_event_distribution_with_Res}
\end{figure}

Figure~\ref{fig:fig_event_distribution_with_Res} shows the zenith angle distribution of $\mu^-$ events
before and after folding in the resolution functions. The angular dependence seems to get only slightly diluted.
This is attributed to the good angular resolution of the detector.

Table~\ref{tab_number_events} shows the total number of muon events with measured energy range 0.8-10.8 GeV at various stages of the analysis for an exposure of 50 kt $\times$ 10 years. Note the sharp fall in statistics due to 
the reconstruction efficiencies. The reconstruction efficiencies are particularly poor for 
the near-horizontal bins where the reconstruction of the muon tracks is very hard.
The small increase in the number of events after applying the energy resolution function 
is due to the spill-over of events from the low-energy part of the spectrum to measured 
energies greater than 0.8 GeV. The spillover to the energy bins with $E_\mu>$10.8 GeV is comparatively small. The zenith angle resolution 
leaves the number of muon events nearly unchanged. 

  \begin{table}[H]

  \begin{center}
  \begin{tabular}{|c|c|c|c|c|}
     
  \hline
   & $\mu^-$ & $\mu^+$ \\
  \hline
 Unoscillated & 14311 & 5723 \\
  \hline
  Oscillated & 10531 & 4188 \\
  \hline
  After Applying Reconstruction and CID Efficiencies & 4941 & 2136 \\
  \hline
  After Applying ($E$, $\cos \theta$) Resolutions & 5270 & 2278 \\
  \hline
  \end{tabular}
  \end{center}

  \caption{Number of muon events produced in CC $\nu_\mu$ interactions at various stages of the analysis for an exposure of 50 kt $\times$ 10 years in the energy range 0.8-10.8 GeV.}
  \label{tab_number_events}

  \end{table}

\subsection{The $\chi^2$ Analysis}

For the $\chi^2$ analysis, we re-bin the data in ten energy bins between $E=0.8-10.8$ GeV with bin width 1 GeV, and twenty $\cos \theta$-bins
between $\cos \theta = -1$ to $+1$, with bin width 0.1. 
The simulated data for ICAL@INO is then statistically treated by defining the following $\chi^2$ :
 \begin{equation}\label{eq_chi2_poisson}
  \chi^{2}_{\rm ino}(\mu^-)
        ={\min_{\xi_k}}  \sum_{i=1}^{N_E} \sum_{j=1}^{N_{\cos \theta}} \left[ 2(N_{ij}^{\rm pred} -  N_{ij}^{\rm obs})
	- 2 N_{ij}^{\rm obs} \: \ln \left( \frac{N_{ij}^{\rm pred}}{N_{ij}^{\rm obs}} \right) \right]
	  + \sum_{k=1}^{5} \xi_{k}^{2}\,,
  \end{equation}
with 
 \begin{equation}
N^{\rm pred}_{ij} = N^{0}_{ij}\bigg(1 + \sum_{k=1}^{5} \pi_{ij}^{k} \xi_{k}\bigg)\,.
\end{equation}
Here $N_{ij}^{\rm pred}$ and $N_{ij}^{\rm obs}$ are the expected and observed number, respectively, of $\mu^-$ events ($N_{\mu^-}^D$) in a given ($E$, $\cos \theta$) bin, while $N_E$ (=10) is the number of energy bins and 
$N_{\cos\theta}$ (=20) is the number of zenith angle bins. 
$N_{ij}^{\rm obs}$ is calculated for a set of assumed ``true value" of the oscillation parameters,  
listed in Table \ref{tab_osc_param_input}. $N_{ij}^0$ is the predicted number of events for a given set of oscillation parameters without the systematic errors included. The systematic uncertainties are included via 
the ``pull" variables $\xi_k$, one each for every systematic uncertainty $\sigma_k$. Here $\pi_{ij}^k$ is the change in the number of events in the $(ij)^{\rm th}$ bin caused by varying the value of $k^{\rm th}$ pull variable $\xi_{k}$ by $\sigma_k$. For determining $\pi_{ij}^k$, we have used a procedure similar to the one described in \cite{GonzalezGarcia:2004wg}.

In this analysis we have considered the following five systematic uncertainties. We take 20\% error on the 
flux normalization, 10\% error on cross sections, and an overall 5\% error on the total number of events. 
In addition, we take a 5\% uncertainty on the zenith angle dependence of the flux, and an 
energy dependent ``tilt error" is included according to the following prescription. 
The event spectrum is calculated with the predicted atmospheric neutrino fluxes and 
then with the flux spectrum shifted
according to
\begin{equation}
\Phi_\delta (E) = \Phi_0 (E) \left ( \frac{E}{E_0} \right )^\delta \simeq \Phi_0(E) \left ( 1+\delta\ln\frac{E}{E_0}
\right ) \,,
\end{equation}
where $E_0=2$ GeV and $\delta$ is the $1\sigma$ systematic tilt error, taken to be 5\%. The difference between
$\Phi_\delta (E)$ and $\Phi_0 (E)$ is then included as the error on the flux. 

For each set of oscillation parameters, 
we calculate the $\chi^2$ separately for the $\mu^-$ and $\mu^+$ data samples, and 
add them to obtain the total $\chi^2$ as
\begin{equation}
\chi^2_{\rm ino} =  \chi^{2}_{\rm ino}(\mu^-) +  \chi^{2}_{\rm ino}(\mu^+)
\,.
\end{equation}

Since in this analysis we are mainly interested in constraining $\theta_{23}$ and 
$|\Delta m^2_{32}|$, and the variation of $\theta_{12}$ or $|\Delta m^2_{21}|$ within the current error bars is observed
not to affect the results, we take the value of these two parameters to be fixed to those given in Table \ref{tab_osc_param_input}.
On the parameter $\sin^22\theta_{13}$, we impose a prior to allow for the uncertainty in its current measurement :
  \begin{equation}\label{eq_chisq_prior}
   \chi^2 = \chi^2_{\rm ino} +
	\left( \frac{\sin^2  2\theta_{13}({\rm true}) - \sin^2 2\theta_{13}}{\sigma_{\sin^2  2\theta_{13}}} \right)^2
	\,,
  \end{equation}
where $\sigma_{\sin^2  2\theta_{13}}$ is the current $1\sigma$ error on $\sin^22\theta_{13}$, and is taken as 0.013 in our analysis.  Of course, during the operation of INO, the error $\sigma_{\sin^2  2\theta_{13}}$ will decrease, and within a few years, $\sin^2 2\theta_{13}$ may be considered to be a fixed parameter.

\section{Precision Measurement of $\sin^2 \theta_{23}$ and $|\Delta m^2_{32}|$}

\begin{figure}
 \centering
\subfigure[]{
\includegraphics[scale=0.38,keepaspectratio=true]{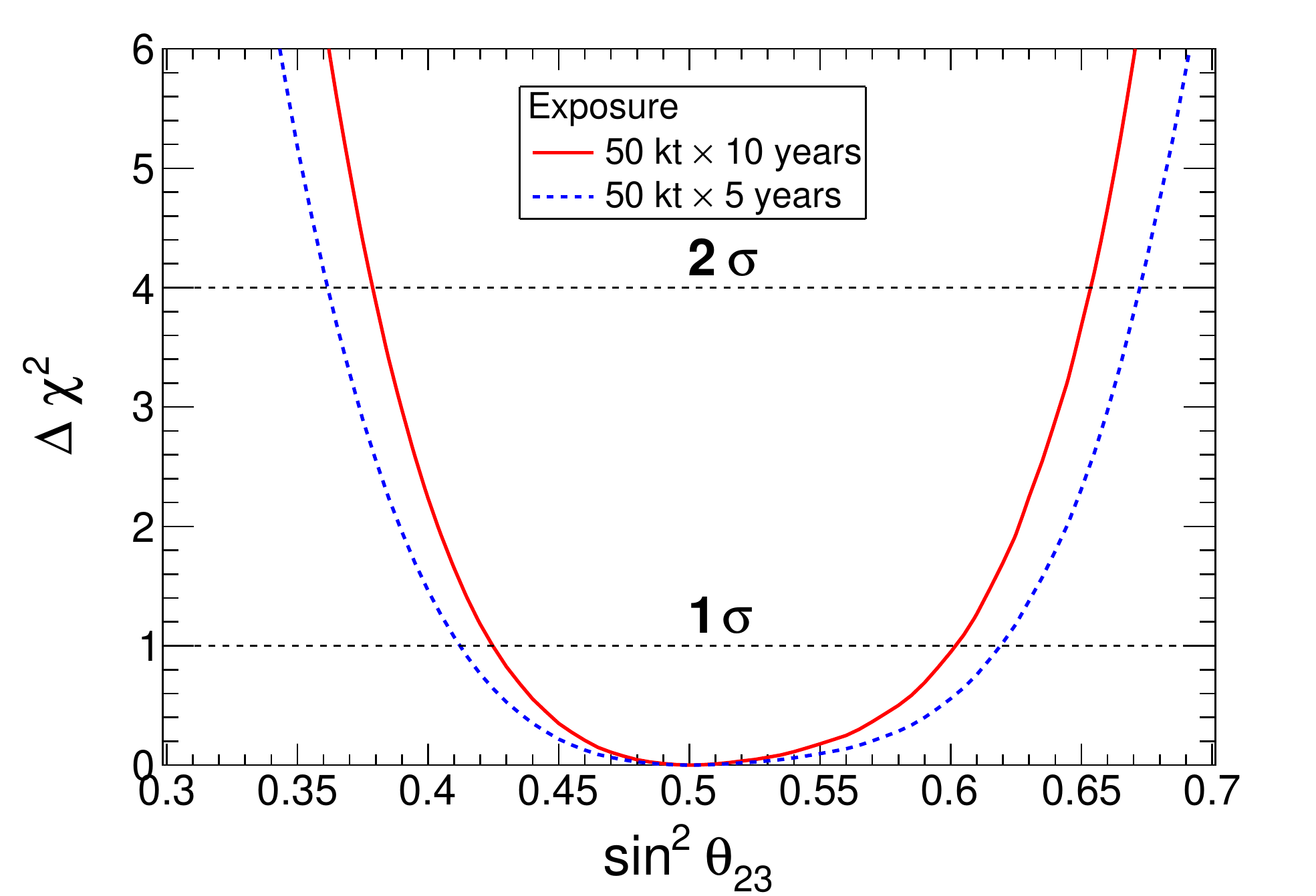}
 \label{fig:fig_chi2_function_th23}
} 
\subfigure[]{
 \includegraphics[scale=0.38,keepaspectratio=true]{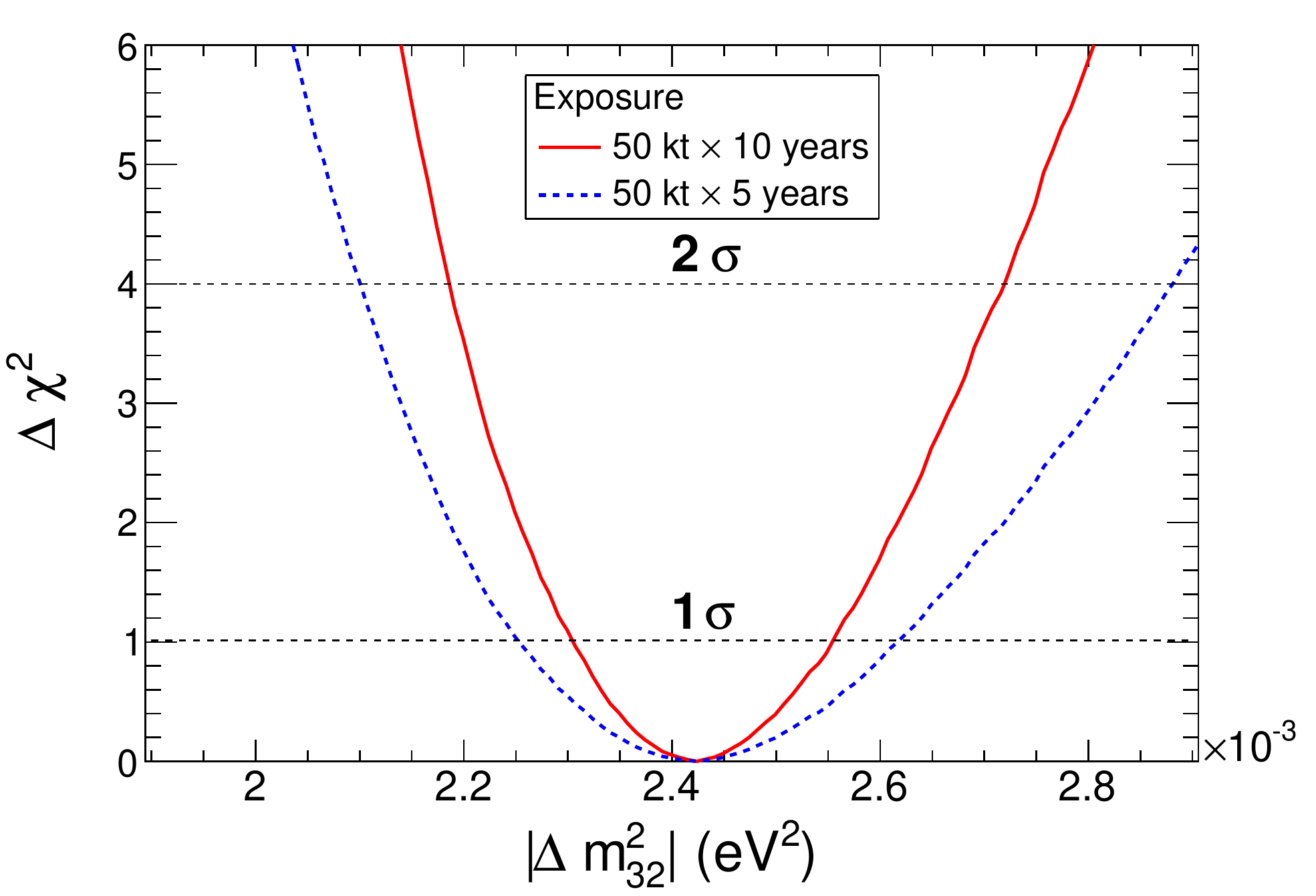}
 \label{fig:fig_chi2_function_dms32}
}
 \caption{The panel (a) shows the $\chi^2$ as a function of $\sin^2 \theta_{23}$ for $|\Delta m^2_{32}|$ = 2.424 $\times$ $10^{-3}$ $\rm eV^2$ and $\sin^2 \theta_{23}(\rm true)=0.5$ . The panel (b) shows the $\chi^2$ as a function of $|\Delta m^2_{32}|$ for  $\sin^2 \theta_{23}$ = 0.5 and $|\Delta m^2_{32}|(\rm true)$ = 2.424 $\times$ $10^{-3}$ $\rm eV^2$}
 \label{fig:fig_chi2_function}
\end{figure}

We start by presenting the reach of the ICAL for the parameters $\sin^2 \theta_{23}$ and $|\Delta m^2_{32}|$ separately. The true values of all parameters are given in Table \ref{tab_osc_param_input}.
Note that we use the parameter $\Delta m^2_{32}$ ( = $\Delta m^2_{31}$ - $\Delta m^2_{21}$) instead of $|\Delta m^2_{31}|$, the current limits on which are given in \cite{Fogli:2012ua}. The $\chi^2$ values as functions of $\sin^2\theta_{23}$ and $|\Delta m^2_{32}|$ are shown in figs.~\ref{fig:fig_chi2_function_th23} and \ref{fig:fig_chi2_function_dms32}, respectively. Note that the minimum value of $\chi^2$ vanishes, since
MC fluctuations in the observed data have been reduced due to the scaling from an exposure of 50 kt $\times$ 1000 years.

The precision on these parameters may be quantified by
\begin{equation}
{\rm precision} = \frac{p_{max}-p_{min}}{p_{max}+p_{min}}
\,,
\end{equation}
where $p_{max}$ and $p_{min}$ are the largest and smallest value of the concerned oscillation 
parameters determined at the given C.L. from the atmospheric neutrino measurements at ICAL for a given exposure.
We find that after 5 years of running of this experiment, ICAL would be able to measure $\sin^2\theta_{23}$ to a precision of $20\%$ and $|\Delta m^2_{32}|$ to 7.4\% at 1$\sigma$. With 10 years exposure, these numbers improve to 17\% and 5.1\% for 
$\sin^2 \theta_{23}$ and $|\Delta m^2_{32}|$, respectively. The precision on  $\sin^2 \theta_{23}$ is mainly governed by the muon reconstruction efficiency and is expected to improve with it. It will also improve as the systematic errors are reduced. If the flux normalization error were to come down from 20\% to 10\%, the precision on $\sin^2 \theta_{23}$ would improve to 14\% for 10 years of exposure. Reducing the zenith angle error from 5\% to 1\% would also improve this precision to $\sim$ 14\%. On the other hand, the precision on $|\Delta m^2_{32}|$ is governed by the ability of the detector to determine the value of $L/E$ for individual events accurately. This depends on the energy- and $\cos \theta$- resolution of the detector. 

A few more detailed observations may be made from the $\chi^2$ plots in fig. \ref{fig:fig_chi2_function}. From fig.~\ref{fig:fig_chi2_function_th23} one can notice that the precision on $\theta_{23}$ when it is in the first octant ($\sin^2 \theta_{23} < 0.5$) is slightly better than 
when it is in the second octant ($\sin^2 \theta_{23} > 0.5$), even though the muon neutrino survival probability depends 
on $\sin^2 2\theta_{23}$ at the leading order. 
This asymmetry about $\sin^2 \theta_{23}=0.5$ stems mainly
from the full three-flavor analysis that we have performed in this study. In particular, we have 
checked that the non-zero value of $\theta_{13}$ is responsible for the asymmetry observed in this figure.
On the other hand, $\chi^2$ asymmetry about the true value of $|\Delta m^2_{32}|$ observed in Fig.~\ref{fig:fig_chi2_function_dms32} is an effect that is present even with a two-flavor analysis. 

\begin{figure}[H]
 \centering
 \subfigure[Precision reach for 5 years run]{
 \includegraphics[scale=0.37,keepaspectratio=true]{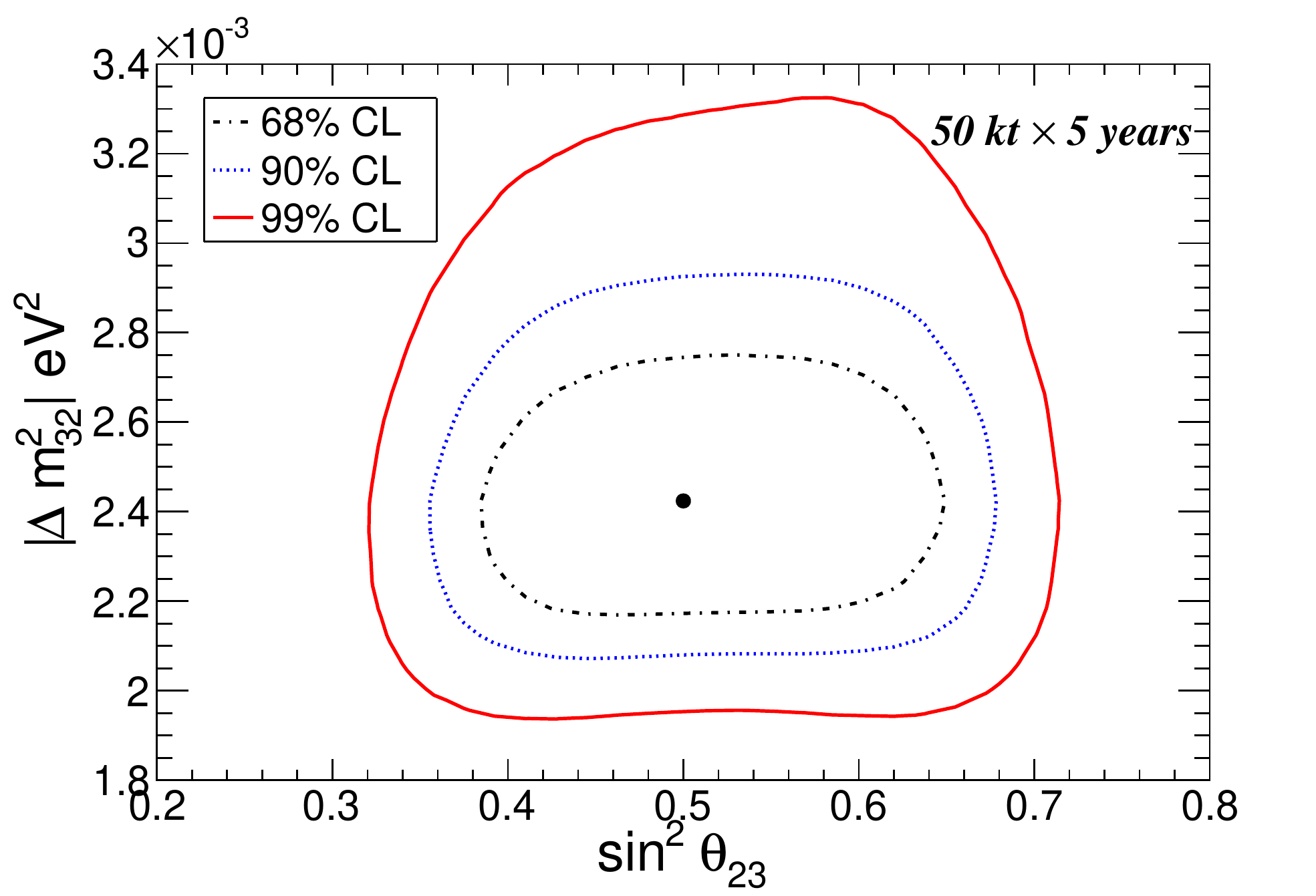}
 \label{fig:5yrs}
 }
  \subfigure[Precision reach for 10 years run]{
 \includegraphics[scale=0.37,keepaspectratio=true]{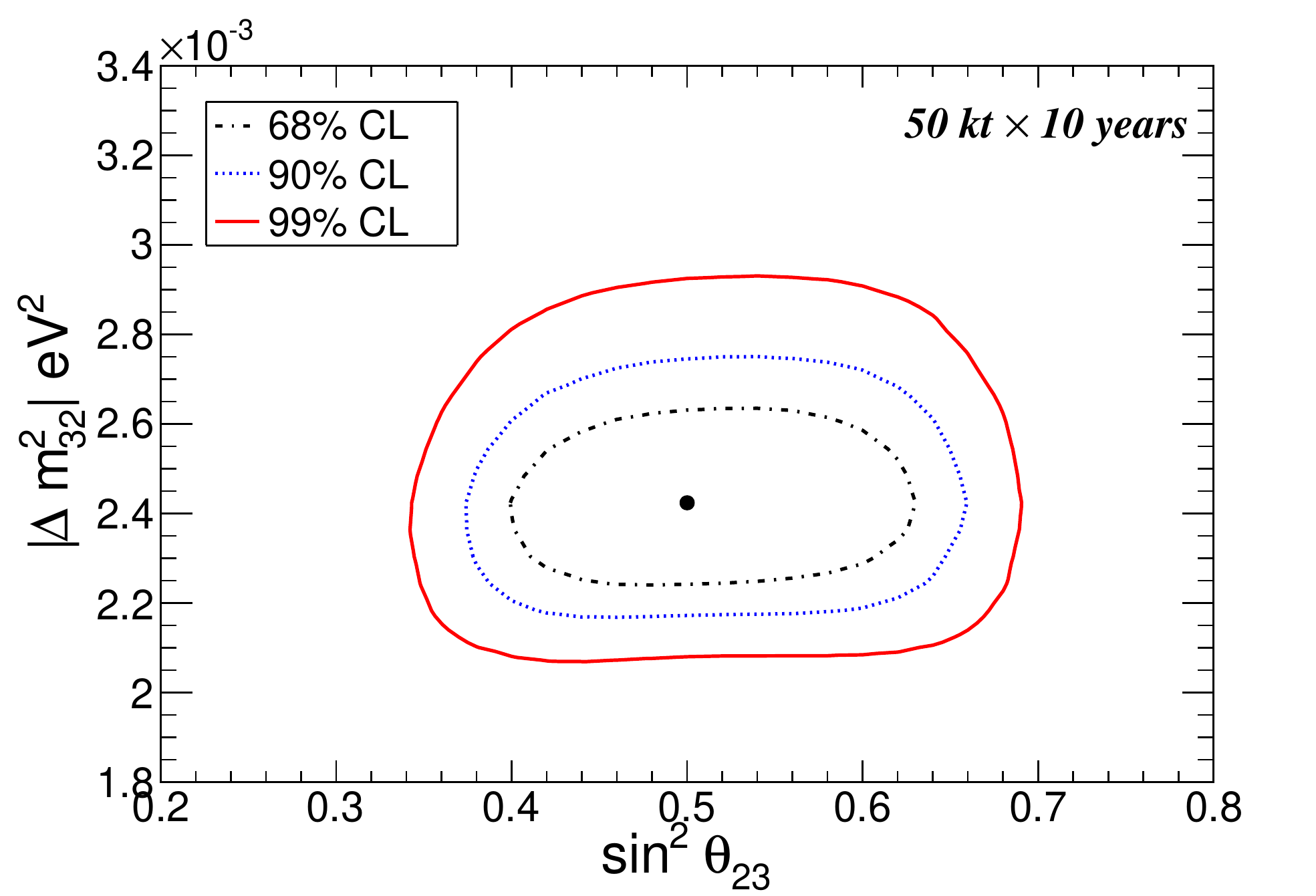}
 \label{fig:10yrs}
 }
 \caption{The precision reach expected at ICAL in the $\sin^2\theta_{23}-|\Delta m^2_{32}|$ plane at various confidence levels. The black(broken), blue(dotted) and red(solid) lines show 68\%, 90\% and 99\% C.L contours.
The true values of $\sin^2\theta_{23}$ and $|\Delta m^2_{32}|$ used for generating data are shown by 
the black dots. The true values of other parameters used are given in Table \ref{tab_osc_param_input}. Panel 
(a) is for five-year running of the 50 kt detector while (b) is for ten years exposure.}
 \label{fig:fig_CL_ssq2th23_1d0}
\end{figure}

The precisions obtainable at the ICAL for $\sin^2 \theta_{23}$ and $|\Delta m^2_{32}|$ are expected to be correlated. We therefore present the correlated reach of ICAL for these parameters in figs.~\ref{fig:5yrs} and \ref{fig:10yrs}. These are the main results of our analysis. As noted above, our three-neutrino analysis should be sensitive to the octant of $\theta_{23}$. Therefore we choose to present our results in terms of $\sin^2 \theta_{23}$ instead of $\sin^2 2\theta_{23}$. Though the constant-$\chi^2$ contours still look rather symmetric about $\sin^2 \theta_{23}=0.5$, that is mainly due to the true value of $\sin^2 \theta_{23}$ being taken to be 0.5. The values of  $\sin^2 \theta_{23}$ away from 0.5 would make the contours asymmetric and would give rise to some sensitivity to the octant of $\theta_{23}$, as we shall see later. \\

\begin{figure}
 \centering
 \includegraphics[scale=0.55,keepaspectratio=true]{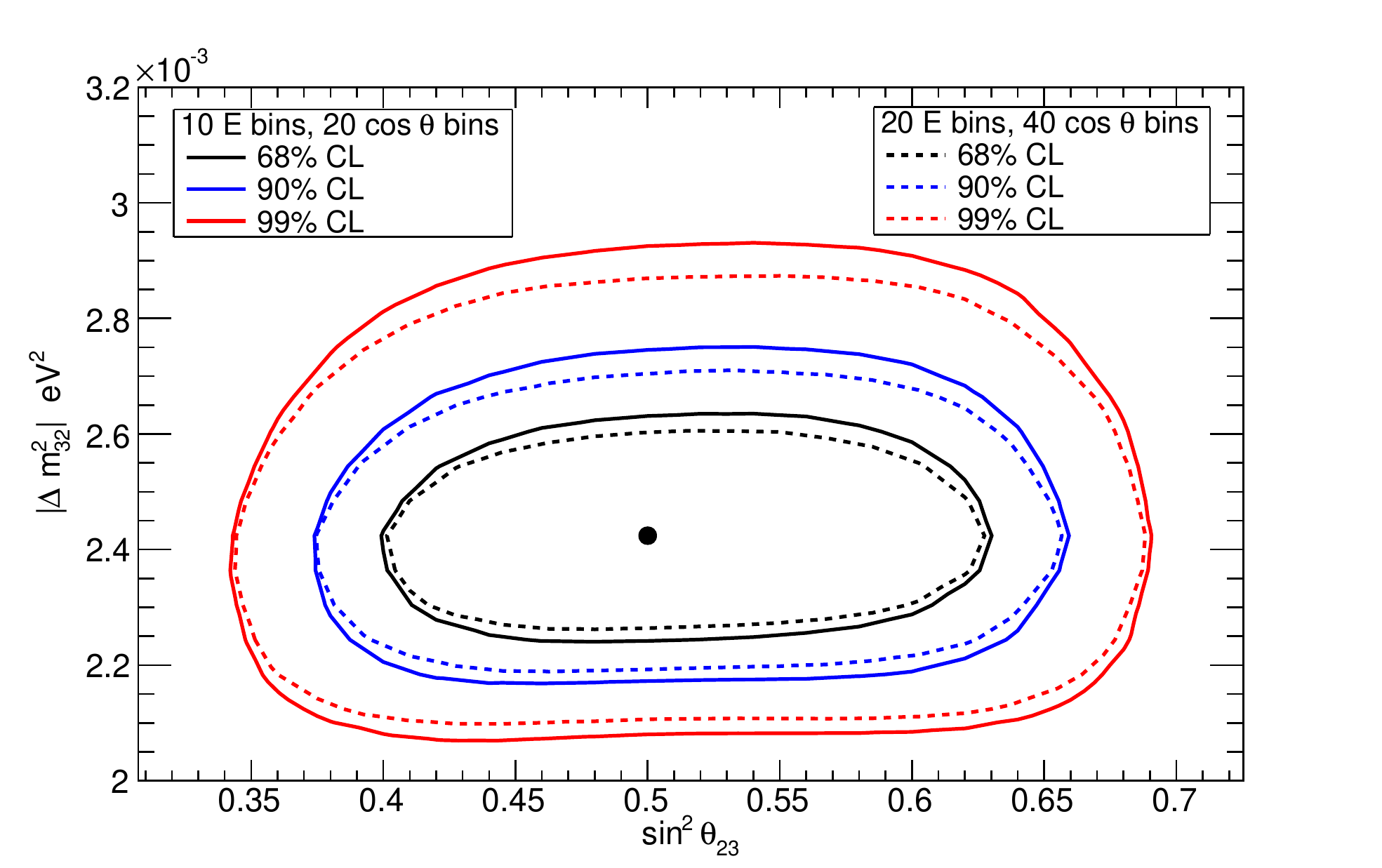}
 \caption{Effect of binning on the precision reach expected from the analysis of 50 kt $\times $ 10 years 
 data of atmospheric neutrinos in ICAL. The black, blue and red lines show 68\%, 90\% and 99\% C.L contours. The solid lines show the contours for (E, $\cos \theta$) bin widths of (1 GeV, 0.1) while the broken lines show the contours for (E, $\cos \theta$) bin widths of (0.5 GeV, 0.05).}
 \label{fig:bins}
 \end{figure}

Note that in our analysis, we have taken $\cos \theta$ bins of width 0.1. The muon angle resolution in ICAL is however better than $1^\circ$ 
for almost all values of the zenith angle. The reason we had to take such 
wide angle bins is because of limited statistics. In order to study the 
impact of our choice of binning on the precision measurements of 
$\sin^2 \theta_{23}$ and $|\Delta m^2_{32}|$ in ICAL, we 
reduce the $\cos \theta$ bin size to 0.05 and the size of the $E$ bins to 0.5 GeV.
In Fig.~\ref{fig:bins}, we show the effect of taking these finer bins in E and $\cos\theta$ on the precision reach of the two parameters. 
The figure corresponds to 10 years of ICAL exposure. We see that the finer bins bring only a marginal improvement 
in the precision measurement of both $\sin^2 \theta_{23}$ and $|\Delta m^2_{32}|$. For further results in this paper, 
we continue to use energy bins of width 1 GeV and zenith angle bins of width 0.1. The optimization with variable bin widths will be a part of future work.

\begin{figure}[H]
 \centering
 \includegraphics[scale=0.7,keepaspectratio=true]{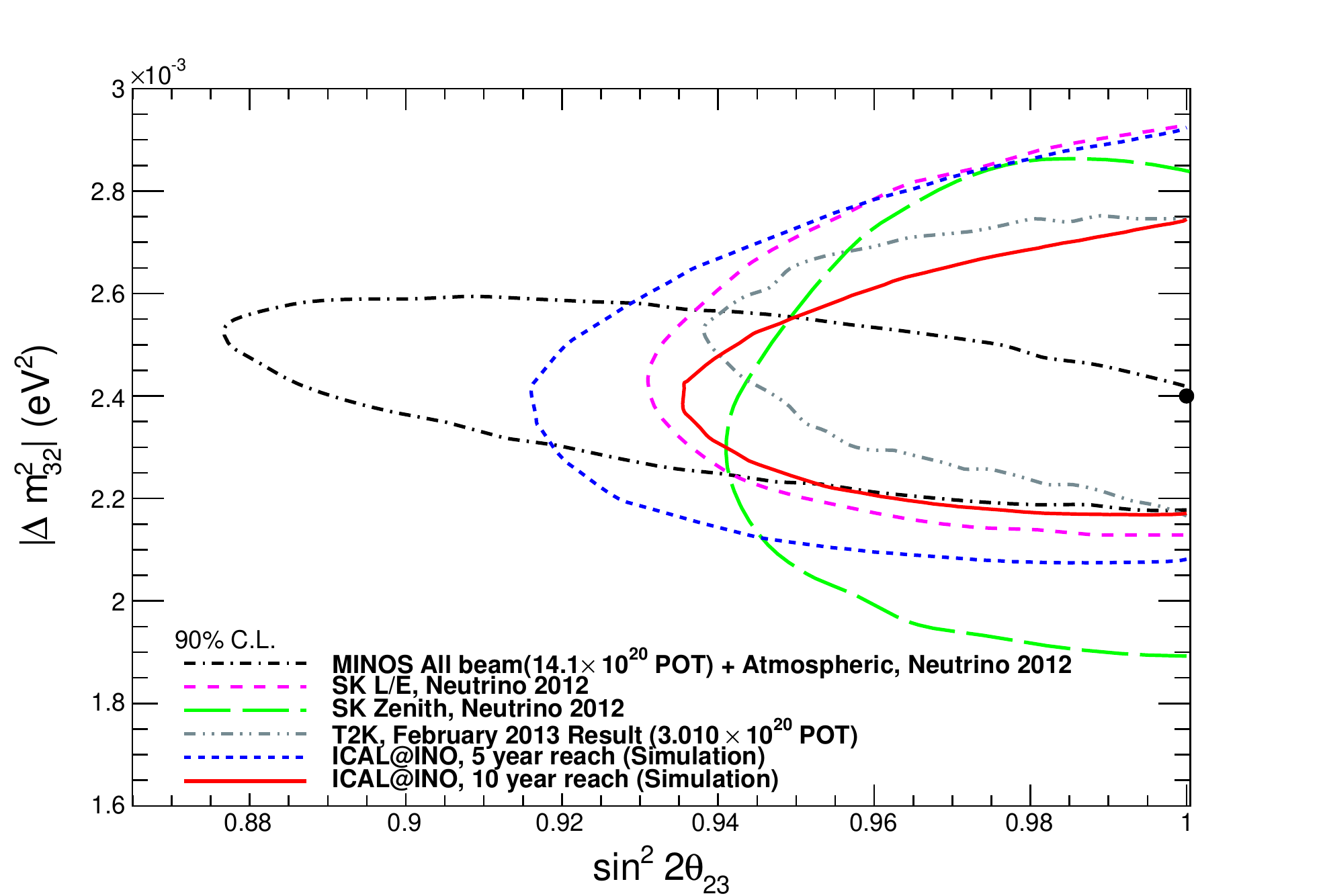}   
 \caption{Comparison of the reach of ICAL@INO with the current results from other neutrino experiments.
 The black dot in the figure denotes the point where the ICAL data was generated. 
 The true values of the other oscillation parameters are given in Table \ref{tab_osc_param_input}.
 }
 \label{fig:fig_CL_Comparision}
\end{figure}

In fig.~\ref{fig:fig_CL_Comparision} we show the comparison of the precision reach on the atmospheric neutrino oscillatino parameters 
at ICAL with that obtained from other experiments currently. Note that here we use the parameter $\sin^2 2\theta_{23}$ instead of $\sin^2 \theta_{23}$ in order to enable a direct comparison.
The blue and red lines show the expected sensitivity from atmospheric neutrino measurements at 
ICAL after 5 years and the 10 years exposure, respectively. The green line is the 90\% C.L.-allowed contour obtained by the zenith angle analysis of SK atmospheric neutrino measurements (shown at the Neutrino 2012 conference), while the pink line is 
the contour obtained by their $L/E$ analysis \cite{sknu2012}. The black line 
shows the 90\% C.L. allowed region given by the combined analysis of the full MINOS data 
including $10.71\times 10^{20}$ POT for the $\nu_\mu$-beam, $3.36\times 10^{20}$ POT for the 
$\bar\nu_\mu$-beam, as well as the atmospheric neutrino data corresponding to an 
exposure of 37.9 kt-years \cite{minosnu2012}. The grey (dot-dot-dashed) line shows the recent T2K $\nu_\mu$ disappearance
analysis results for $3.01 \times 10^{20}$ POT \cite{t2k2013_Feb}.

From fig.~\ref{fig:fig_CL_Comparision} it may be observed that, with 5 years of exposure, ICAL will be able to 
almost match the precision on $|\Delta m^2_{32}|$ obtained from the SK $L/E$ analysis currently.
With 10 years data this will improve, though it will still not be comparable to 
the precision we already have from the MINOS experiment. 
Since the direction of neutrinos in MINOS is known accurately, their $L/E$ is known to a greater precision and their measurement of 
$|\Delta m^2_{32}|$ is consequently more accurate. 
The precision of ICAL on $\sin^2 2\theta_{23}$ in 10 years may be expected to be comparable to what we currently have from SK.
(Of course by the time ICAL completes 10 years, SK would have gathered even more data.)
This precision is controlled to a large extent by the total number of events.
We can see that the sensitivity of ICAL to $\sin^2 2\theta_{23}$ and  $|\Delta m^2_{32}|$
is not expected to surpass the precision we already have from the current set of 
experiments. In fact, the precision on these parameters are expected to improve 
significantly with the expected data from T2K \cite{Abe:2011ks,Abe:2012gx} and NO$\nu$A \cite{Ayres:2004js}, and ICAL will not
be competing with them as far as these precision measurements are concerned. (The recent T2K results \cite{t2k2013_Feb} already claim a precision comparable to ICAL reach in 10 years.) The ICAL data will however give complementary information on these parameters, which will significantly contribute to the improvement of the precision on the global fit.

\begin{figure}[H]
 \centering
\subfigure[$\sin^2 2\theta_{23}$ = 0.9, first octant ($\sin^2 \theta_{23}=0.342$)]
{
 \includegraphics[scale=0.37,keepaspectratio=true]{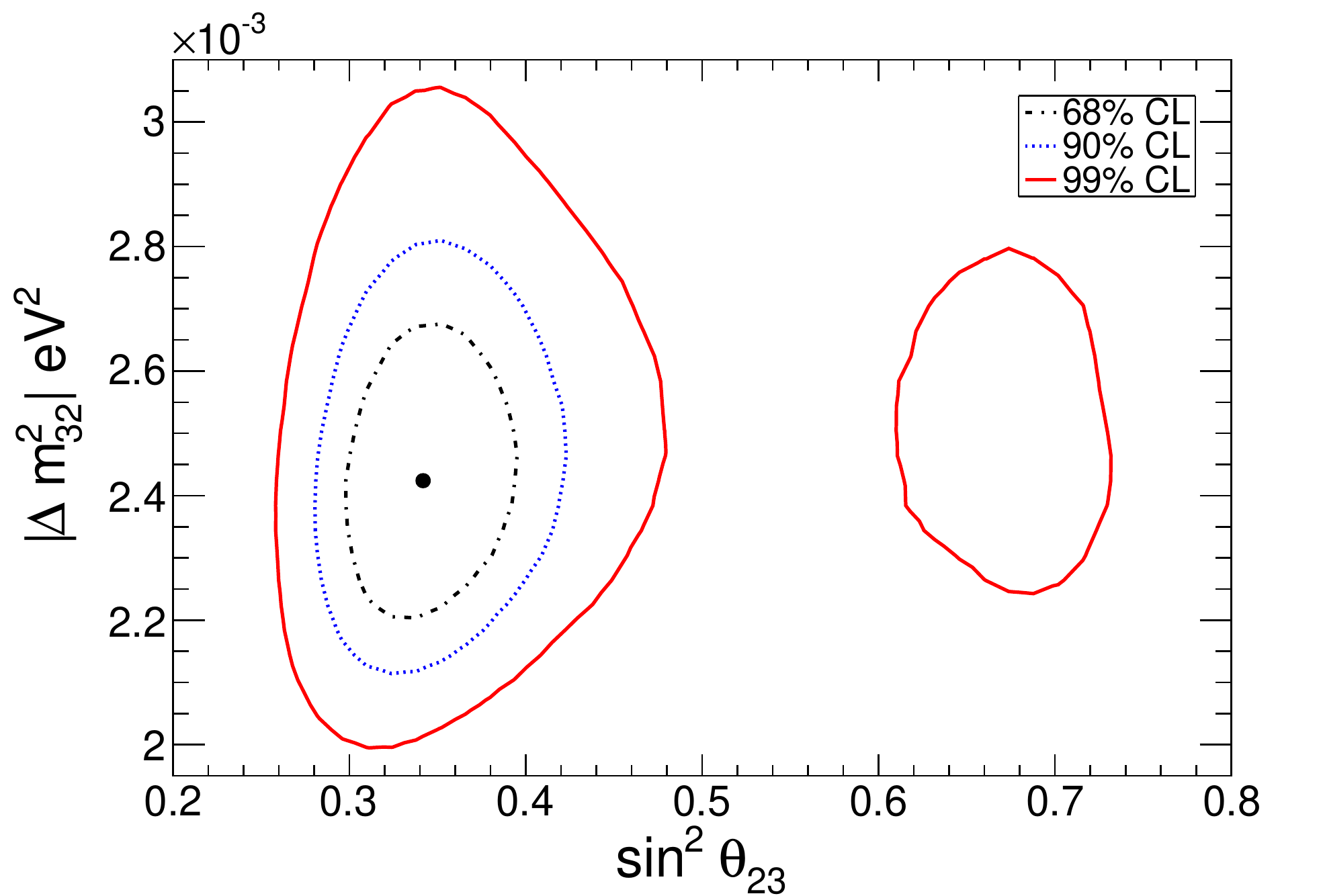}
 \label{fig:fig_CL_ssqth23_0d90_1}
} \quad
\subfigure[$\sin^2 2\theta_{23}$ = 0.9, second octant ($\sin^2 \theta_{23}=0.658$)]
{
\includegraphics[scale=0.37,keepaspectratio=true]{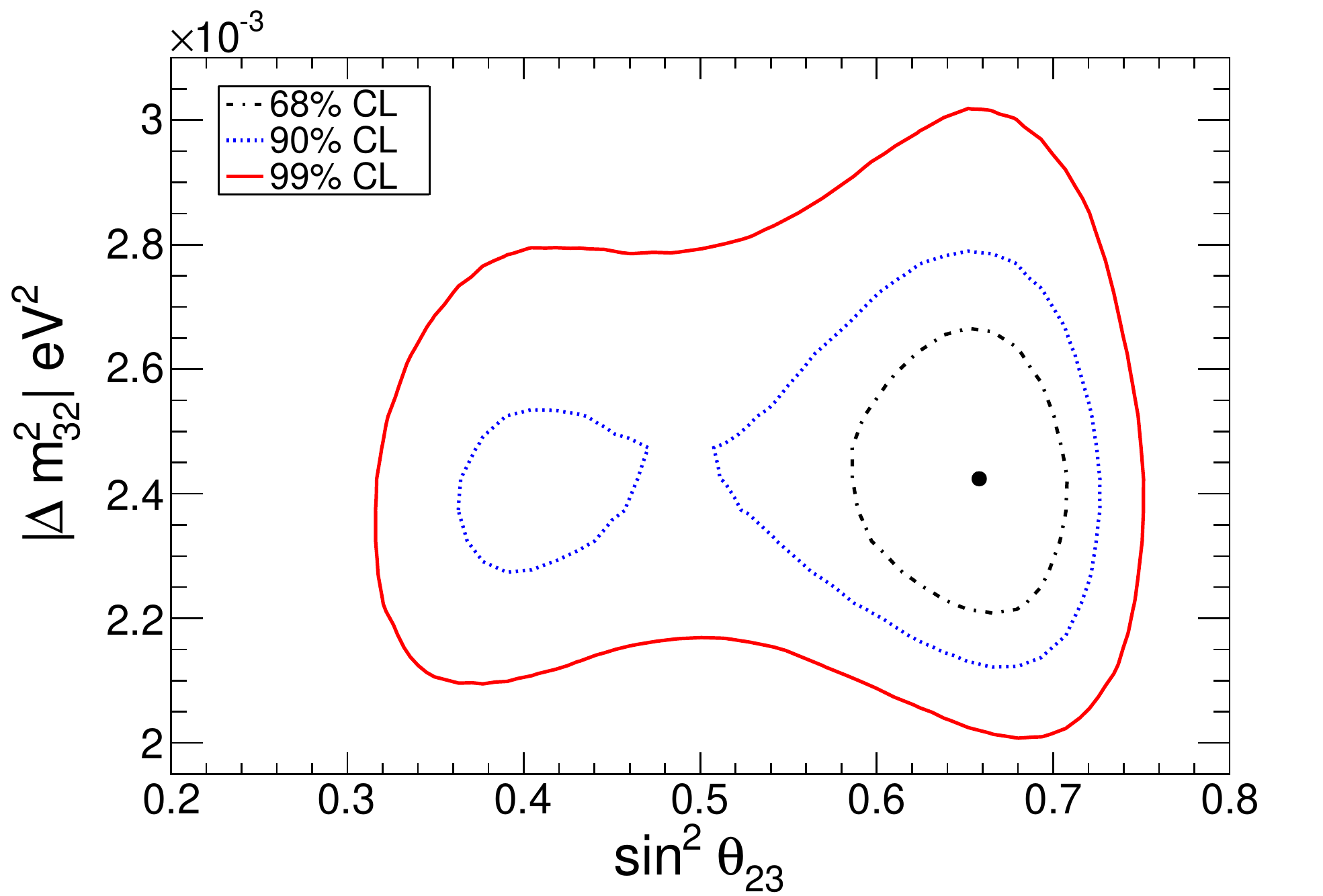}
\label{fig:fig_CL_ssqth23_0d90_2}
}
\subfigure[$\sin^2 2\theta_{23}$ = 0.95, first octant ($\sin^2 \theta_{23}=0.388$)]{
 \includegraphics[scale=0.37,keepaspectratio=true]{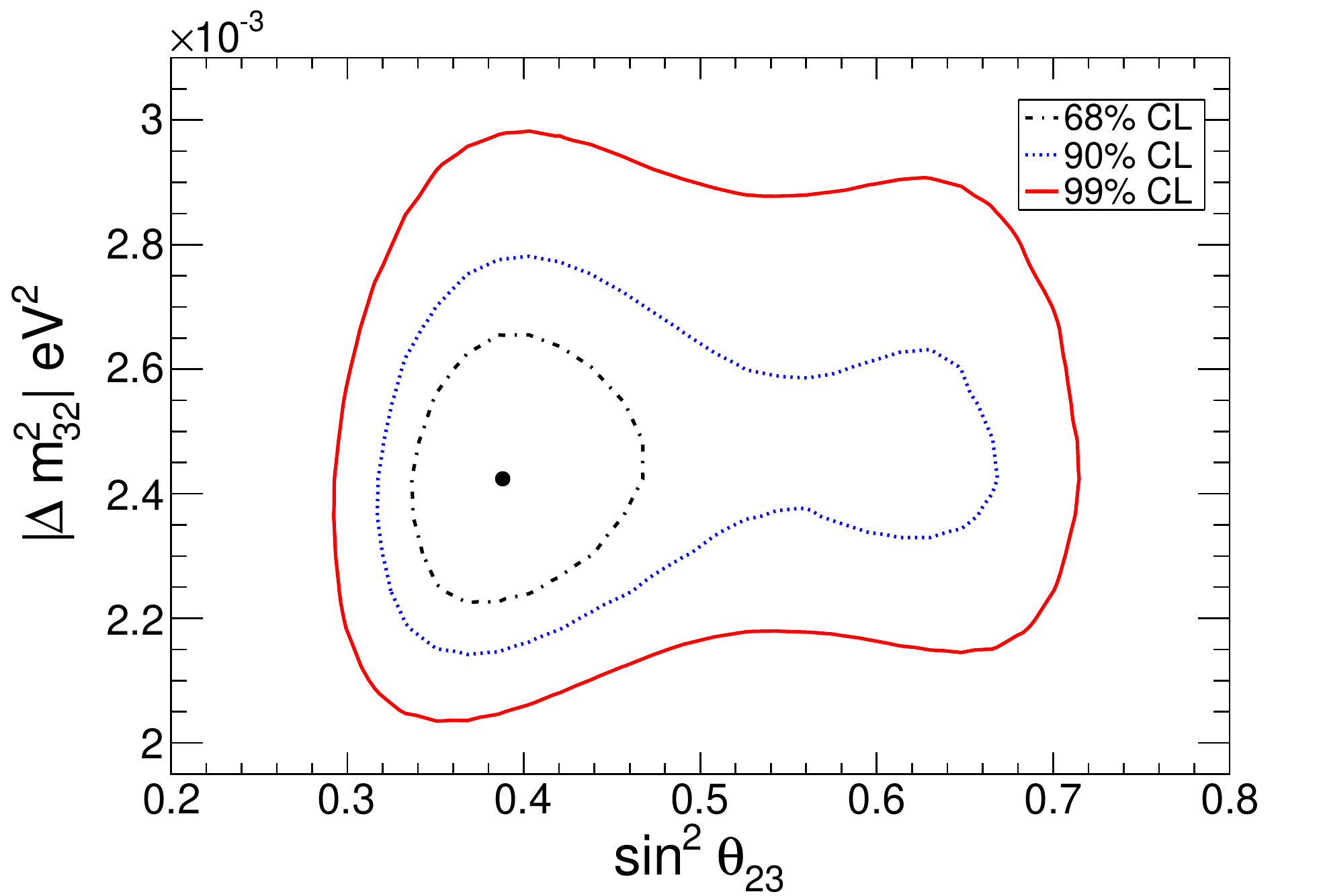}
 \label{fig:fig_CL_ssqth23_0d95_1}
} \quad
\subfigure[$\sin^2 2\theta_{23}$ = 0.95, second octant ($\sin^2  \theta_{23}=0.612$)]{
\includegraphics[scale=0.37,keepaspectratio=true]{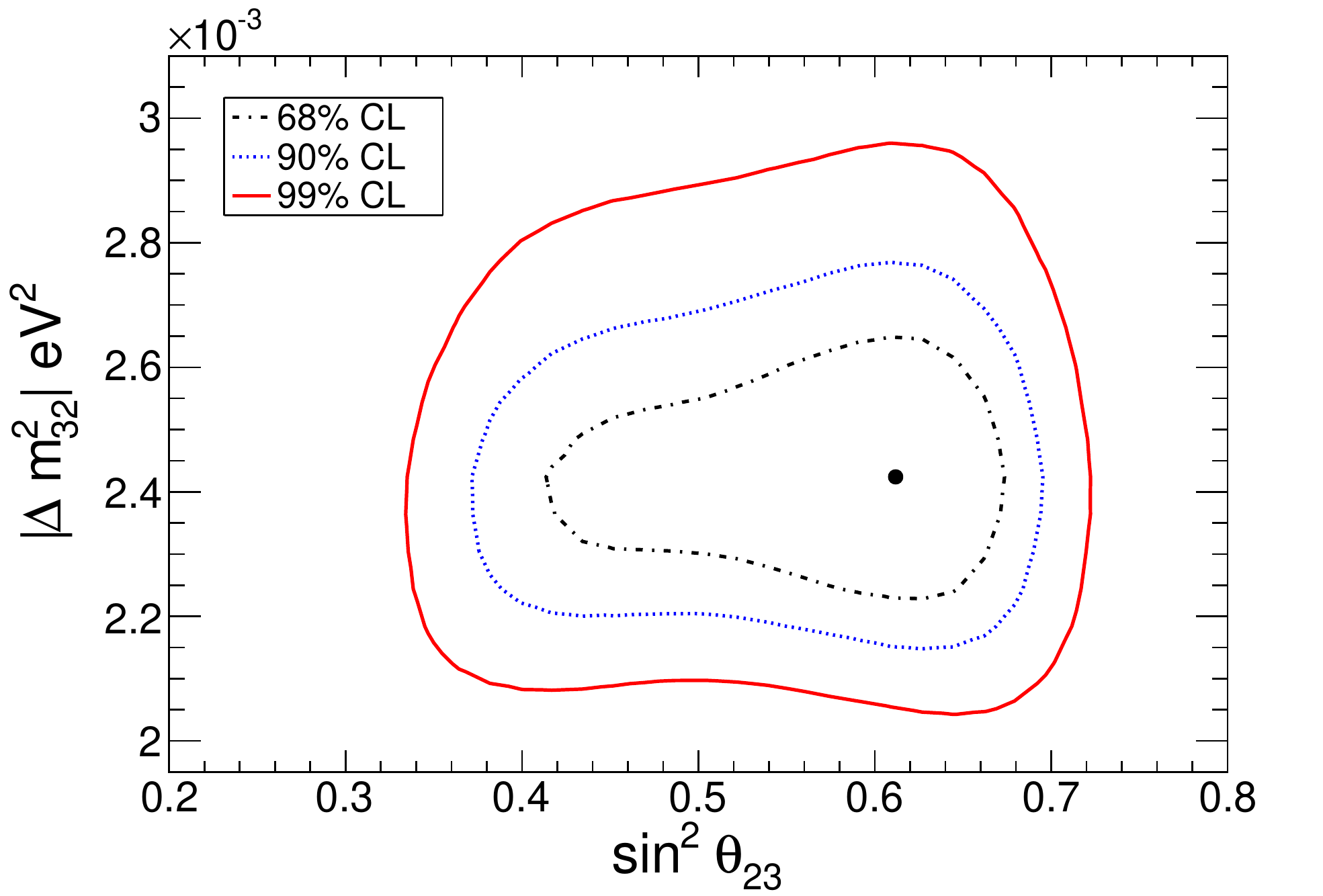}
\label{fig:fig_CL_ssqth23_0d95_2}
}
\caption{The projected reach in the $\sin^2\theta_{23}-|\Delta m^2_{32}|$ plane for 
four different non-maximal choices of $\theta_{23}$. The black(broken), blue(dotted) and red(solid) lines show 68\%, 90\% and 99\% C.L. contours for 10 years of 50 kt ICAL run. Note that we use normal hierarchy, and assume that it is already known.}
 \label{fig:fig_CL_octant_sensitivity}
\end{figure}

Earth matter effects in atmospheric neutrinos can be used to 
distinguish maximal from non-maximal $\theta_{23}$ mixing and can lead to the determination 
of the correct $\theta_{23}$ octant \cite{Choubey:2005zy,Barger:2012fx,Indumathi:2006gr}.  
We show in Fig.~\ref{fig:fig_CL_octant_sensitivity}
the potential of 10 years of ICAL run for distinguishing a non-maximal value of $\theta_{23}$ from maximal mixing in the case where $\sin^2 2\theta_{23}=0.90$ ($\sin^2 \theta_{23}$ = 0.342, 0.658) and $\sin^2 2\theta_{23}$ = 0.95 ($\sin^2 \theta_{23}$ = 0.388, 0.612). 
Note that the current 3$\sigma$ allowed range of $\sin^2 2\theta_{23}$ is (0.91, 1.0). The figure shows that, if the value of $\theta_{23}$ is near the current 3$\sigma$ bound and in the first octant, then it may be possible to exclude maximal mixing to 99\% C.L. with this 2-parameter analysis. If $\theta_{23}$ is in the second octant, or if $\sin^2 2\theta_{23}$ is larger than 0.9, the exclusion of the maximal mixing becomes a much harder task.

Fig.~\ref{fig:fig_CL_octant_sensitivity} can also be used to quantify the reach of ICAL for determining the correct octant of $\theta_{23}$, if the value of $\sin^2 2\theta_{23}$ is known. This can be seen by comparing the $\chi^2$ value corresponding to the true value of $\sin^2 \theta_{23}$, but in the wrong octant, with that corresponding to the true value of $\sin^2 \theta_{23}$. We find that, for $\sin^2 2\theta_{23} = 0.9$, i.e. just at the allowed 3$\sigma$ bound, the octant can be identified at $>$95\% C.L.  with 10 years of ICAL run if $\theta_{23}$ is in the first octant. However if $\theta_{23}$ is in the second octant, the identification of the octant would be much harder: $\theta_{23}$ in the wrong octant can be disfavored only to about 85\% C.L.. The situation is more pessimistic if $\sin^2 2\theta_{23}$ is closer to unity.

The precision on $|\Delta m^2_{32}|$ will keep improving with ongoing and future long baseline experiments. The inclusion of the information
 may improve the chance of ICAL@INO being able to identify deviation of $\theta_{23}$ from maximal mixing and its octant to some extent, however that exercise is beyond the scope of this paper.
 
\section{Summary}

In the analysis, we have calculated the projected reach of the ICAL experiment at INO for precise determination of the atmospheric 
neutrino parameters. We have used NUANCE simulated data on muon energies and directions using the resolutions and efficiencies obtained from a complete detector simulation by the INO collaboration. We incorporate the oscillation physics using the re-weighting method and calculate the values of $\chi^2$ in the $\sin^2 \theta_{23}$--$|\Delta m^2_{32}|$ plane using the pull method that takes into account systematic errors on (i) flux normalization (ii) cross sections (iii) neutrino flux tilt (iv) zenith angle dependence of flux and (v) overall systematic error. 

We present uncorrelated as well as correlated constraints on the value of  $\sin^2 \theta_{23}$ and $|\Delta m^2_{32}|$ expected to be obtained after 10 years of running of the 50 kt ICAL. We find that the values of $\sin^2 \theta_{23}$ and $|\Delta m^2_{32}|$ may be determined at an accuracy of 17\% and 5.1\% respectively. The sensitivities with the data at ICAL only are not expected to be better than what we already have, indeed some of the other experiments in the next decade may do much better. However the measurement at ICAL will be complementary and may be expected to contribute significantly towards the precision of parameters in a global fit. 

As far as the sensitivity to the $\theta_{23}$ octant and its deviation from maximality is concerned, we find that 10 years of ICAL can exclude maximal mixing or the $\theta_{23}$ value in the other octant to $>$95\% C.L. only if the actual $\theta_{23}$ is in the first octant and close to the current 3$\sigma$ lower bound to 99 \% C.L.. Indeed, the octant identification seem to be beyond the reach of any single experiment in the next decade.

This paper presents the first study on the reach of ICAL@INO for  precision of atmospheric neutrino parameters,
using the complete detector simulation.
Note that in this analysis we have used only the information on muon events.
However, the ICAL atmospheric neutrino experiment will also record and measure the hadrons 
associated with the charged current interaction of the muon-type neutrinos. Inclusion of this data-set 
into the analysis is expected to provide energy and angle reconstruction of the neutrino. This could 
lead to an improved sensitivity of the detector to the oscillation parameters. The analysis including the 
hadrons along with the muons is a part of the ongoing effort of the INO-ICAL collaboration. 
In addition, the updates and improvements in the muon momentum reconstruction algorithm as well as the optimization of our analysis procedure are likely to improve the results presented in this paper.

\section{Acknowledgemets}
This work is a part of the ongoing effort of INO-ICAL collaboration to study various physics potential of
the proposed INO-ICAL detector. Many members of the collaboration have contributed for the completion
of this work. We are very grateful to G. Majumder and A. Redij for their developmental work on the ICAL detector simulation package.
We also thank A. Chatterjee, K. Meghna, K. Rawat, D. Indumathi for their contributions in characterizing the muon detector response of the ICAL detector. In addition, we thank N. Mondal, D. Indumathi, S. Uma Sankar, S. Goswami, N. Sinha, P. Ghoshal and M. Naimuddin for intensive discussions on the oscillation analysis in regular meetings. T.T. thanks Costas Andreopoulos for discussions on the neutrino event generators.  A.D. and S.C. acknowledge partial support from the European Union FP7 ITN INVISIBLES (Marie Curie Actions,PITN-GA-2011-289442). Finally we thank V. Datar, K. Kar and A. Raychaudhuri for critical reading of the manuscript.


\begin{thebibliography}{9}

\bibitem{Wendell:2010md} 
  R.~Wendell {\it et al.}  [Super-Kamiokande Collaboration],
  Phys.\ Rev.\ D {\bf 81}, 092004 (2010)
  [arXiv:1002.3471 [hep-ex]].

\bibitem{Aharmim:2011vm} 
  B.~Aharmim {\it et al.}  [SNO Collaboration],
  arXiv:1109.0763 [nucl-ex].

\bibitem{Abe:2008aa} 
  S.~Abe {\it et al.}  [KamLAND Collaboration],
  Phys.\ Rev.\ Lett.\  {\bf 100}, 221803 (2008)
  [arXiv:0801.4589 [hep-ex]].


\bibitem{An:2012eh} 
  F.~P.~An {\it et al.}  [DAYA-BAY Collaboration],
  Phys.\ Rev.\ Lett.\  {\bf 108}, 171803 (2012)
  [arXiv:1203.1669 [hep-ex]].

\bibitem{Ahn:2012nd} 
  J.~K.~Ahn {\it et al.}  [RENO Collaboration],
  Phys.\ Rev.\ Lett.\  {\bf 108}, 191802 (2012)
  [arXiv:1204.0626 [hep-ex]].

\bibitem{Abe:2012tg} 
  Y.~Abe {\it et al.}  [Double Chooz Collaboration],
  Phys.\ Rev.\ D {\bf 86}, 052008 (2012)
  [arXiv:1207.6632 [hep-ex]].

\bibitem{Ahn:2006zza} 
  M.~H.~Ahn {\it et al.}  [K2K Collaboration],
  Phys.\ Rev.\ D {\bf 74}, 072003 (2006)
  [hep-ex/0606032].

\bibitem{Adamson:2012gt} 
  P.~Adamson {\it et al.}  [MINOS Collaboration],
  Phys.\ Rev.\ D {\bf 86}, 052007 (2012)
  [arXiv:1208.2915 [hep-ex]].

\bibitem{Abe:2011ks} 
  K.~Abe {\it et al.}  [T2K Collaboration],
  Nucl.\ Instrum.\ Meth.\ A {\bf 659}, 106 (2011)
  [arXiv:1106.1238 [physics.ins-det]].

\bibitem{Pontecorvo:1967fh} 
  B.~Pontecorvo,
  Sov.\ Phys.\ JETP {\bf 26}, 984 (1968)
  [Zh.\ Eksp.\ Teor.\ Fiz.\  {\bf 53}, 1717 (1967)].

\bibitem{Maki:1962mu} 
  Z.~Maki, M.~Nakagawa and S.~Sakata,
  Prog.\ Theor.\ Phys.\  {\bf 28}, 870 (1962).

\bibitem{Apollonio:2002gd} 
  M.~Apollonio {\it et al.}  [CHOOZ Collaboration],
  Eur.\ Phys.\ J.\ C {\bf 27}, 331 (2003)
  [hep-ex/0301017].

\bibitem{Fogli:2012ua} 
  G.~L.~Fogli, E.~Lisi, A.~Marrone, D.~Montanino, A.~Palazzo and A.~M.~Rotunno,
  Phys.\ Rev.\ D {\bf 86}, 013012 (2012)
  [arXiv:1205.5254 [hep-ph]].

\bibitem{Tortola:2012te} 
  D.~V.~Forero, M.~Tortola and J.~W.~F.~Valle,
  Phys.\ Rev.\ D {\bf 86}, 073012 (2012)
  [arXiv:1205.4018 [hep-ph]].

\bibitem{Athar:2006yb} 
  M.~S.~Athar {\it et al.}  [INO Collaboration],
  INO-2006-01.

\bibitem{Abe:2011ts} 
  K.~Abe, T.~Abe, H.~Aihara, Y.~Fukuda, Y.~Hayato, K.~Huang, A.~K.~Ichikawa and M.~Ikeda {\it et al.},
  arXiv:1109.3262 [hep-ex].

\bibitem{Rubbia:2004nf} 
  A.~Rubbia,
  Nucl.\ Phys.\ Proc.\ Suppl.\  {\bf 147}, 103 (2005)
  [hep-ph/0412230].

\bibitem{Akiri:2011dv} 
  T.~Akiri {\it et al.}  [LBNE Collaboration],
  arXiv:1110.6249 [hep-ex].

\bibitem{Koskinen:2011zz} 
  D.~J.~Koskinen,
  Mod.\ Phys.\ Lett.\ A {\bf 26}, 2899 (2011).

\bibitem{Akhmedov:2012ah} 
  E.~K.~.Akhmedov, S.~Razzaque and A.~Y.~.Smirnov,
  JHEP {\bf 02}, 082 (2013)
  [JHEP {\bf 1302}, 082 (2013)]
  [arXiv:1205.7071 [hep-ph]].

\bibitem{Ayres:2004js} 
  D.~S.~Ayres {\it et al.}  [NOvA Collaboration],
  hep-ex/0503053.

\bibitem{Spergel:2003cb} 
  D.~N.~Spergel {\it et al.}  [WMAP Collaboration],
  Astrophys.\ J.\ Suppl.\  {\bf 148}, 175 (2003)
  [astro-ph/0302209].

\bibitem{Hannestad:2010kz} 
  S.~Hannestad,
  Prog.\ Part.\ Nucl.\ Phys.\  {\bf 65}, 185 (2010)
  [arXiv:1007.0658 [hep-ph]].

\bibitem{Lesgourgues:2012uu} 
  J.~Lesgourgues and S.~Pastor,
  Adv.\ High Energy Phys.\  {\bf 2012}, 608515 (2012)
  [arXiv:1212.6154 [hep-ph]].

\bibitem{KlapdorKleingrothaus:2000sn} 
  H.~V.~Klapdor-Kleingrothaus, A.~Dietz, L.~Baudis, G.~Heusser, I.~V.~Krivosheina, S.~Kolb, B.~Majorovits and H.~Pas {\it et al.},
  Eur.\ Phys.\ J.\ A {\bf 12}, 147 (2001)
  [hep-ph/0103062].

\bibitem{Ackerman:2011gz} 
  N.~Ackerman {\it et al.}  [EXO-200 Collaboration],
  Phys.\ Rev.\ Lett.\  {\bf 107}, 212501 (2011)
  [arXiv:1108.4193 [nucl-ex]].

\bibitem{Kraus:2004zw} 
  C.~.Kraus, B.~Bornschein, L.~Bornschein, J.~Bonn, B.~Flatt, A.~Kovalik, B.~Ostrick and E.~W.~Otten {\it et al.},
  Eur.\ Phys.\ J.\ C {\bf 40}, 447 (2005)
  [hep-ex/0412056].

\bibitem{Wolf:2008hf} 
  J.~Wolf [KATRIN Collaboration],
  Nucl.\ Instrum.\ Meth.\ A {\bf 623}, 442 (2010)
  [arXiv:0810.3281 [physics.ins-det]].
  
\bibitem{Ghosh:2012px} 
  A.~Ghosh, T.~Thakore and S.~Choubey,
  JHEP {\bf 1304}, 009 (2013)
  [arXiv:1212.1305 [hep-ph]].
  

\bibitem{Reference_Muon}
    ``Simulation study of the sensitivity of the ICAL detector to muons'',
    INO Collaboration,
    under preparation.

\bibitem{Reference_Hadron}
    ``Hadron energy response of the ICAL detector at INO'',
    M.M. Devi, A. Ghosh, D. Kaur, L.S. Mohan, S. Choubey et al.,
    INO Collaboration,
    [arXiv:1304.5115 [hep-ex]].

\bibitem{Samanta:2008af} 
  A.~Samanta,
  Phys.\ Rev.\ D {\bf 80}, 113003 (2009)
  [arXiv:0812.4639 [hep-ph]].

\bibitem{Samanta:2010xm} 
  A.~Samanta and A.~Y.~.Smirnov,
  JHEP {\bf 1107}, 048 (2011)
  [arXiv:1012.0360 [hep-ph]].

\bibitem{Agostinelli:2002hh} 
  S.~Agostinelli {\it et al.}  [GEANT4 Collaboration],
  Nucl.\ Instrum.\ Meth.\ A {\bf 506}, 250 (2003).

\bibitem{Allison:2006ve} 
  J.~Allison, K.~Amako, J.~Apostolakis, H.~Araujo, P.~A.~Dubois, M.~Asai, G.~Barrand and R.~Capra {\it et al.},
  IEEE Trans.\ Nucl.\ Sci.\  {\bf 53}, 270 (2006).

\bibitem{Casper:2002sd} 
  D.~Casper,
  Nucl.\ Phys.\ Proc.\ Suppl.\  {\bf 112}, 161 (2002)
  [hep-ph/0208030].

\bibitem{Honda:2011nf} 
  M.~Honda, T.~Kajita, K.~Kasahara and S.~Midorikawa,
  Phys.\ Rev.\ D {\bf 83}, 123001 (2011)
  [arXiv:1102.2688 [astro-ph.HE]].

\bibitem{Barger:1980tf} 
  V.~D.~Barger, K.~Whisnant, S.~Pakvasa and R.~J.~N.~Phillips,
  Phys.\ Rev.\ D {\bf 22}, 2718 (1980).

\bibitem{GonzalezGarcia:2004wg} 
  M.~C.~Gonzalez-Garcia and M.~Maltoni,
  Phys.\ Rev.\ D {\bf 70}, 033010 (2004)
  [hep-ph/0404085].

\bibitem{sknu2012}
	Y. Itow,
	SK Collaboration,
	``SK results presented at Neutrino 2012'',
	http://neu2012.kek.jp.
	
\bibitem{minosnu2012}
	P. Adamson et al.,
	MINOS Collaboration,
	``Talk presented at Neutrino 2012, http://www-numi.fnal.gov/pr\_plots/CC\_Brag\_Slide\_Jun2012\_mod1.pdf'',
	[arXiv:1304.6335 [hep-ex]].

\bibitem{t2k2013_Feb}
	T. Dealtry,
	T2K Collaboration,
        ``Muon neutrino disappearance at T2K'',
	http://indico.cern.ch/contributionDisplay.py?contribId=36\&sessionId=12\&confId=214998.
	

\bibitem{Abe:2012gx} 
  K.~Abe {\it et al.}  [T2K Collaboration],
  Phys.\ Rev.\ D {\bf 85}, 031103 (2012)
  [arXiv:1201.1386 [hep-ex]].

\bibitem{Choubey:2005zy} 
  S.~Choubey and P.~Roy,
  Phys.\ Rev.\ D {\bf 73}, 013006 (2006)
  [hep-ph/0509197].

\bibitem{Barger:2012fx} 
  V.~Barger, R.~Gandhi, P.~Ghoshal, S.~Goswami, D.~Marfatia, S.~Prakash, S.~K.~Raut and S U.~Sankar,
  Phys.\ Rev.\ Lett.\  {\bf 109}, 091801 (2012)
  [arXiv:1203.6012 [hep-ph]].

\bibitem{Indumathi:2006gr} 
  D.~Indumathi, M.~V.~N.~Murthy, G.~Rajasekaran and N.~Sinha,
  Phys.\ Rev.\ D {\bf 74}, 053004 (2006)
  [hep-ph/0603264].

\end{thebibliography}
\end{document}